\documentclass[10pt]{IEEEtran}
\usepackage{array}
\usepackage[switch,mathlines]{lineno}
\usepackage{soul}
\usepackage{graphicx}
\usepackage{cite}
\usepackage[cmex10]{amsmath}
\usepackage{algorithm}

\usepackage{algpseudocode}
\usepackage{fixltx2e}
\usepackage{CJK}
\usepackage[cmex10]{amsmath}
\usepackage{bm}
\usepackage{color}
\usepackage{amsmath,amsthm,amssymb,amsfonts}

\usepackage{arydshln} 

\usepackage{soul}
\usepackage{tikz}
\usepackage{booktabs}
\usepackage{tabularx}

\usepackage{cite}
\usepackage{hyperref}
\hypersetup{
     colorlinks   = true,
     linkcolor    = blue,
     citecolor    = red
}

\usepackage{xfrac}

\makeatletter
\newcommand*{\transpose}{%
  {\mathpalette\@transpose{}}%
}
\newcommand*{\@transpose}[2]{%
  \raisebox{\depth}{$\m@th#1\intercal$}%
}
\makeatother

\usepackage{scalerel}
\usepackage{tikz}
\usetikzlibrary{svg.path}

\usepackage{scalerel}
\usepackage{tikz}
\usetikzlibrary{svg.path}

\definecolor{orcidlogocol}{HTML}{A6CE39}
\tikzset{
    orcidlogo/.pic={
        \fill[orcidlogocol] svg{M256,128c0,70.7-57.3,128-128,128C57.3,256,0,198.7,0,128C0,57.3,57.3,0,128,0C198.7,0,256,57.3,256,128z};
        \fill[white] svg{M86.3,186.2H70.9V79.1h15.4v48.4V186.2z}
        svg{M108.9,79.1h41.6c39.6,0,57,28.3,57,53.6c0,27.5-21.5,53.6-56.8,53.6h-41.8V79.1z M124.3,172.4h24.5c34.9,0,42.9-26.5,42.9-39.7c0-21.5-13.7-39.7-43.7-39.7h-23.7V172.4z}
        svg{M88.7,56.8c0,5.5-4.5,10.1-10.1,10.1c-5.6,0-10.1-4.6-10.1-10.1c0-5.6,4.5-10.1,10.1-10.1C84.2,46.7,88.7,51.3,88.7,56.8z};
    }
}

\newcommand\orcidicon[1]{\href{https://orcid.org/#1}{\mbox{\scalerel*{
                \begin{tikzpicture}[yscale=-1,transform shape]
                \pic{orcidlogo};
                \end{tikzpicture}
            }{|}}}}

\begin{document}
\title{\huge{Stochastic Multi-Agent-Based Model to Measure Community Resilience-Part 2: Simulation Results}}

\author{Jaber Valinejad, Student member, IEEE, Lamine Mili, Life Fellow, IEEE, Konstantinos Triantis, Michael von Spakovsky, C. Natalie van der Wal
\vspace{-1cm}

\thanks{\hspace{-8pt}\underline{~~~~~~~~~~~~~~~~~~~~~~~~~~~~~~~~~~~~~~~~~~~~~~~~~~~}}
\thanks{The work presented in this paper was funded by the National Science Foundation (NSF) under Grant No. 1917308.
J. Valinejad and L. Mili are with the Bradley Department of Electrical and Computer Engineering, Virginia Tech, Northern Virginia Center, Greater Washington D.C., VA 22043, USA (email:{JaberValinejad,lmili}@vt.edu).

 K. P. Triantis is with the Grado Department of Industrial and Systems Engineering, Virginia Tech, Greater Washington D.C., VA 22043-2311 USA (e-mail: triantis@vt.edu).
 
Michael von Spakovsky is with the  Department of Mechanical Engineering, Virginia Tech, Blacksburg, VA 24061
 USA (e-mail:vonspako@vt.edu ).
 
Natalie van der Wal is with the University of Leeds, Leeds University Business School, Center for Decision Research, Leeds, United Kingdom.(e-mail: c.n.vanderwal@leeds.ac.uk)

}
}
\maketitle

\begin{abstract}

In this paper we investigate the resiliency planning of interdependent electric power systems and emergency services. We investigate the effect of the level of empathy, cooperation, coordination, flexibility, and experience of individuals on their mental well-being. Furthermore, we explore the impact of the information that is provided by emergency services and the impact of the availability of electric energy on the physical, mental, and social well-being of individuals. For our simulations, we use a stochastic, multi-agent-based numerical framework that is reported in the companion paper for estimating the social well-being of a community when facing natural disasters such as hurricanes, floods, earthquakes, and tsunamis. The performance of the proposed method is assessed by measuring community resilience  for a multitude of effects in the context of two case studies. These effects are analyzed for Gaussian social random characteristics. Each case study considers nine agents, namely, three areas of three communities each, yielding a total of six communities. The results show that a high level of cooperation can positively change individual behavior. In addition, the  relationship  among  the  individuals  of a community is so vital that the society with less population and more empathy may be more resilient than the community with more population and less empathy.

\end{abstract}


\vspace{-0.6cm}

\section*{Nomenclature}
For the notation used throughout this paper, the reader is referred to the companion paper \cite{jaber2019b}.


\section{Introduction}
Emergency preparedness is important for saving lives during disasters. Electric power systems and emergency services need to be resilient during disasters such as hurricanes, floods, earthquakes, and tsunamis. In this paper, we investigate the resiliency planning of interdependent electric power systems and emergency services. To do so, the effect of the level of empathy, cooperation, coordination, flexibility, and experience of individuals on their mental well-being is examined. This is accomplished by applying the framework elaborated in the companion paper\cite{jaber2019b} on community resilience to two case studies. The proposed stochastic multi-agent-based framework allows a multitude of natural disasters and scenarios to be simulated. The goal of these simulations is to gain a better understanding of resiliency so that the critical factors influencing it can be identified and mitigated and lives saved. 

The main characteristic of community resilience is social well-being, which has a mental and physical aspect \cite{links2018copewell, harms2018resilience,cinderby2016building}. Mental well-being includes the following interrelated aspects: emotion, risk perception, information-seeking behavior, compassionate empathy, cooperation, coordination, flexibility, and experience \cite{paperasli,8667341,7637014,stranges2014major, 7782466}. Physical well-being, on the other hand, is affected by the following aspects: an injury factor incurred during a disaster, the availability of emergency services, and the electric energy provided by utilities and distributed energy resources (DERs) \cite{kalt2019conceptualizing,welsch2014electricity, carret2009inappropriate, nikic2009place,thomson2017health}.  

In this paper, two case studies are simulated to understand (1) individual effects on community resilience and (2) the effects of emergency services information and electric energy availability on community resilience. The paper is organized as follows. Section II provides the data and results associated with case study 1.This case study consists of a small community of three socially separated areas, each formed by three empathetic agents\footnote{Each agent represents an individual person. The terms: agent, person, and individual have the same meaning throughput the paper}. In section III, the results for a society with six different communities is presented. It is envisioned that each community possesses a different size population with different social characteristics. Practical ideas like time banking to enhance community resilience are discussed in section IV, while section V ends with some conclusions and future research directions. \par

\section{Method}
In this paper, a stochastic multi-agent-based model is proposed to measure and analyse community resilience. Modeling human-related characteristics, the availability of critical infrastructures during a  disaster,  electricity sharing, and social behavior diffusion is entwined with uncertainties. The uncertainties related to human-community features assumed to follow a Gaussian distribution. In this paper, the following effects are simulated: 
\begin{itemize}
\item The effect of flexibility on human responses to disaster; 
\item The effect of the human experience on the collective behavior and the mental well-being of a society during and after a disaster;
\item The effect of cooperation, coordination, and experience on community resilience to disasters;
\item  The effect of the availability of electric energy and
the willingness to share among the individuals of a community on physical well-being and community resilience to disasters;
\item  The effect of the actions taken by the emergency services, the injury factor of a disaster, and the news polarity \footnote{News polarity measures the degree to which the news are positive or negative.} on the physical and mental well-being of a society;
\item  The effect of varying mass media trends on community resilience to disasters;
\item The effect of the level of compassionate empathy among people on their social well-being;
\item The effect of the occurrence of one disaster in a community on its collective behavior;
\item The effect of the occurrence of two concurrent disasters in two different communities on the human response;
\item The effect of emergencies that arise at different times on community resilience to disasters;
\item The effect of a diversified community population on its social well-being during and after a disaster;
\item An analysis of the effect of the international Emergency Events Database (EM-DAT) on mental and physical well-being as well as community resilience;
\item A discussion about the effect of the time-banking as an alternative currency on  community resilience to disasters; \\
\end{itemize}

The proposed stochastic multi-agent-based model is implemented in two different case studies. This model is verified by The soft validation and sensitivity analyses. The social effect of human characteristics, mass media, and critical infrastructures on community resilience are analyzed in the first case study, i.e., a community of nine persons facing a hurricane. The aim of the second case study, i.e., a society of six separate communities, is to clarify the social effect of different community characteristics and various scenarios of disasters (in terms of time, place, and a specific type of disaster) on mental well-being, physical well-being, and community resilience.

\section{Simulation Results for Case Study 1: Community of Nine Agents Facing a Hurricane}

\begin{figure}
\centering
\includegraphics[width=0.8 \columnwidth]{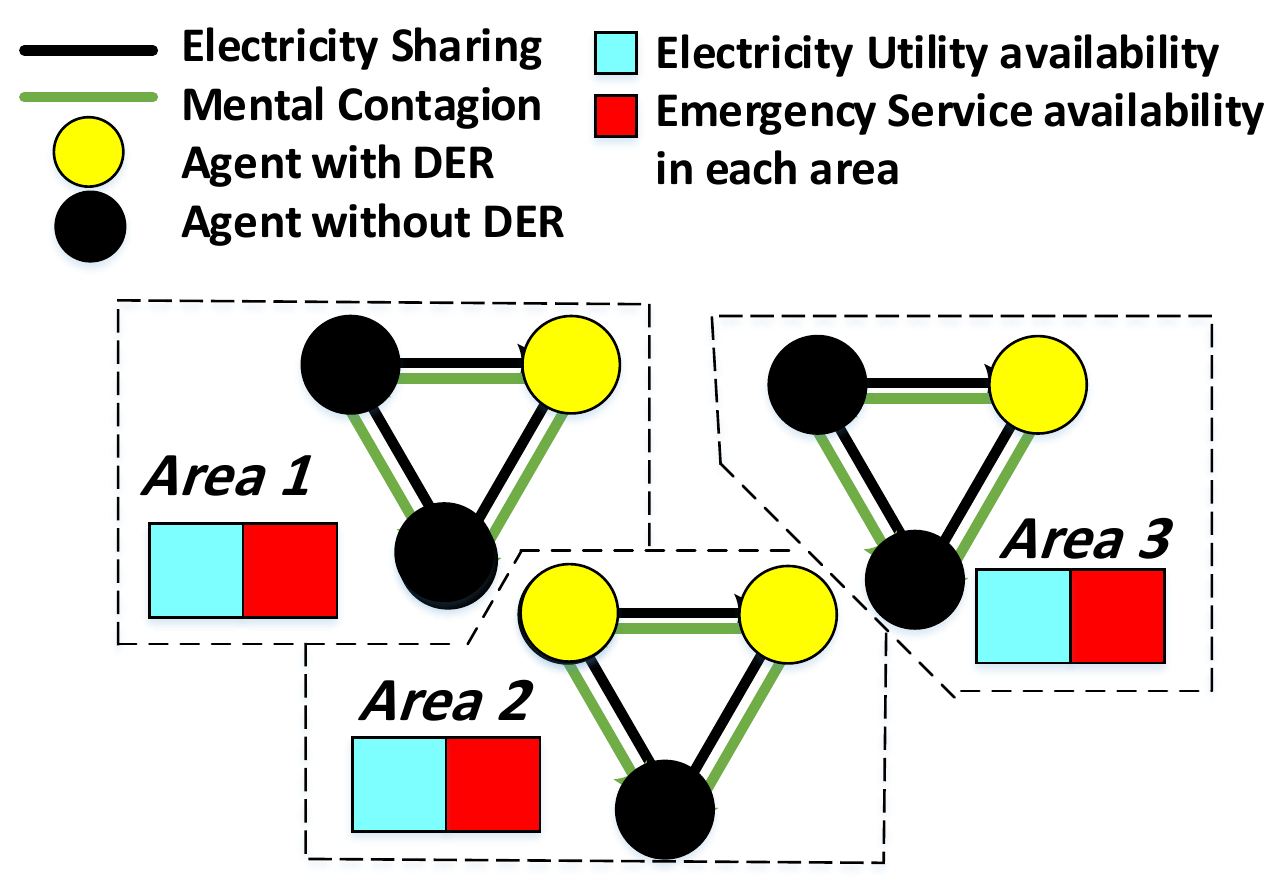}
\setlength{\abovecaptionskip}{-7pt}
\caption{Case study 1: nine agents (prosumers or consumers) during the disaster. Prosumers have accessibility to distributed energy resources. It is assumed that the agents in each area have similar initial behaviors and conditions. There is empathy among individuals in each area, while individuals are not empathetic to individuals in other areas.}
\vspace{-0.4cm}
\label{fig:Fig_343}
\end{figure}

This section analyzes the performance of the proposed dynamical model of a community of nine agents experiencing a hurricane. Specifically, the dynamic changes of the mental and physical characteristics of the agents is assessed.   \\
The first case study aims to clarify the effect of each human feature, including emotion, risk perception, information-seeking behavior, empathy, cooperation, flexibility, and experience on the collective behavior and the mental well-being of society during and after a disaster. In addition, the social effect of mass media, availability of electricity, and the availability of emergency services on community resilience is analyzed. \\  This community, which consists of three areas, is represented in Figure ~\ref{fig:Fig_343}. Each area involves three individuals  empathetic to each other. The individuals of each area do not have any communication with those of another area.\par
The parameter setting of the models of the mental and physical characteristics, mass media, emergency services, and electric grid are provided in Table~\ref{tab:FC0}. These features are assumed to taking a value in the interval [0 1] \cite{paperasli,natalie_13}. The meaning of each value is comprehensively discussed in \cite{jaber2019b}. The electricity consumption of each individual is assumed to be 1 kWh of which 0.8 kWh is supplied by utilities through distribution power lines and 0.2 kWh is supplied by distributed energy resources (DERs), photo-voltaics (PVs), and wind turbines to name a few. 
Furthermore, the fraction of electricity that the DERs supply for each individual $W^{DER}$ is set to 0.2. 

\begin{table}[H]

\caption{Parameter settings for the mental and physical characteristic, mass media, emergency services, and electric grid}
\centering
\scriptsize
\begin{tabular}{|p{0.7 cm}|p{7 cm}| p{0.3 cm}|}
\hline  
Parameter & Definition & Value \\ \hline
$M^{E}_{ti}$&  the level of fear of an individual& 0.5    \\ \hline                      
$M^{R}_{ti}$ & The level of risk perception &  0.5 \\ \hline
$M^{B}_{ti}$& The level of information-seeking behavior& 0.5 \\ \hline
$M^{C}_{ti}$ &The level of  cooperation  & 0.5 \\ \hline
$M^{O}_{ti}$ & The level of personal characteristics& 0.5 \\ \hline
$M^{L}_{ti}$ & The level of experience & 0.5 \\ \hline
$\gamma^{E}_{ij}$ & The level of compassionate empathy between two individuals &  1\\ \hline
$M^{F}_{ti}$ & The level of flexibility & 1\\ \hline
$P_{ti}$ & The level of physical health & 1\\ \hline
$N_{t}$ & The fraction of the event-related information of the public news provided by the mass media (e.g., television, newspapers, social networks) &1\\ \hline
$N^{+}_{t}$ & The fraction of the information conveyed by the mass media that are positive & 0 \\ \hline
$Z_{ti}$ & The level of injury factor of disaster & 1\\ \hline
$Q^{DER}_{ti}$& The fraction of electricity that is available from  DERs to an individual& 1 \\ \hline
$Q^{e}_{ti}$ & The fraction of electricity that is available from utilities to a costumer & 0.5 \\ \hline
$Q^{s}_{t}$ & The degree of help that an individual gets from emergency services during, and after a disaster & 1 \\ \hline
\end{tabular}
\vspace{-0.4cm}
\label{tab:FC0}
\end{table}

\vspace{-0.3cm}
\subsection{Soft Validation of the proposed stochastic multi-agent-based modelling}
At this step, a soft validation is done. Our computational model is verified by Case Study 1 that is taken from \cite{natalie_13}. For this step, information-seeking behavior, the emotion of fear, and bias are considered in the model.  After soft validation, the model is extended to consider the mental resilient-related characteristics, the physical well-being of agents, and critical infrastructures, including emergency services and the power grid. We investigate if the patterns/social phenomena can be simulated with the proposed model. After verification, we pinpoint and analyze the emergent effects that result from the social interactions using multi-agent-based modeling.

\vspace{-0.3cm}
\subsection{Effects of Flexibility on Human Responses}

\begin{figure}
\centering
\includegraphics[width=1 \columnwidth]{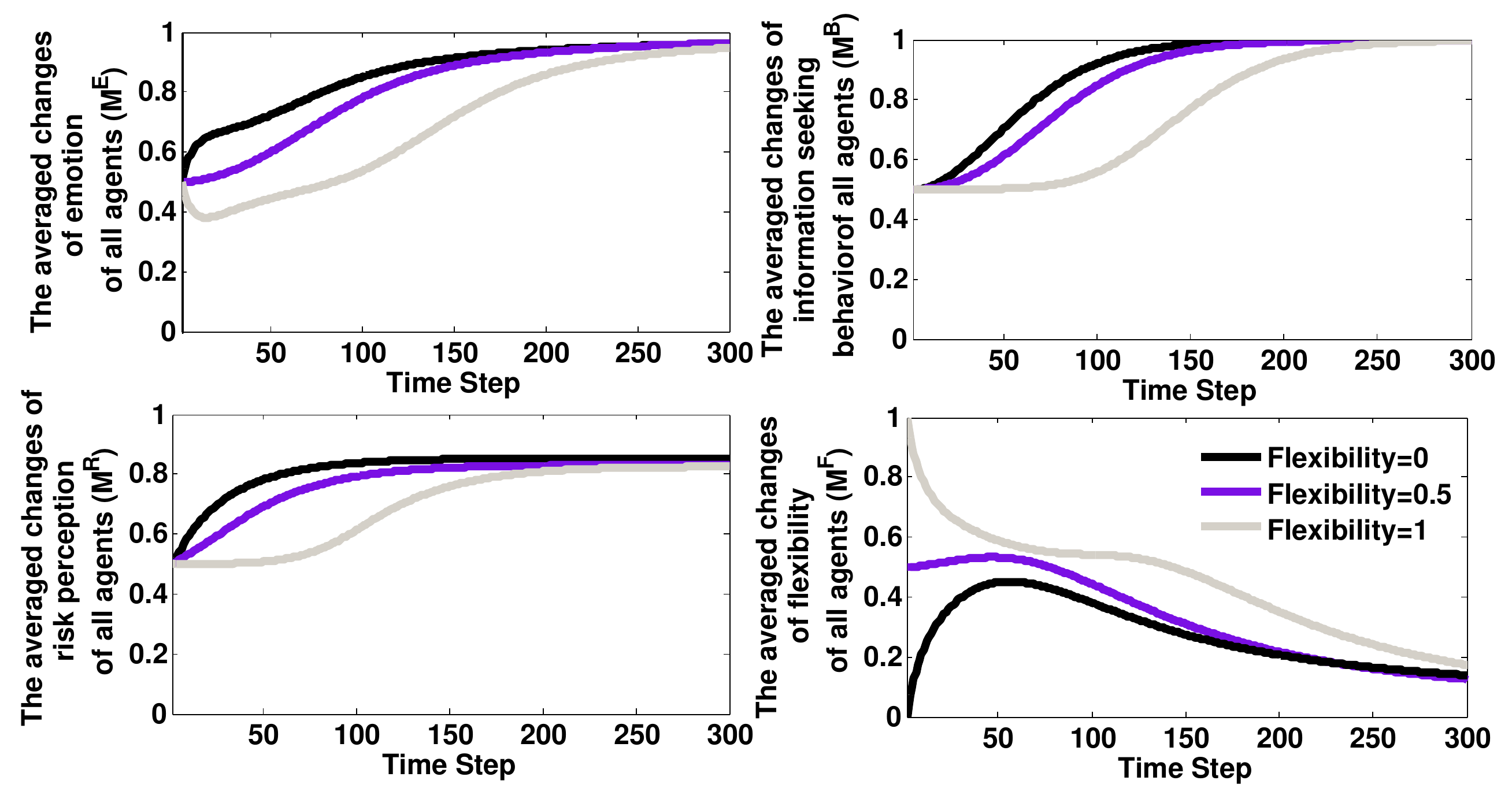}
\setlength{\abovecaptionskip}{-15pt}
\caption{Effects of flexibility on collective behavior and mental characteristics. The dynamic change of emotion, information-seeking behavior, risk perception, and flexibility of all agents are shown. Results are provided for three different initial values of flexibility (0, 0.5, and 1). The time duration of the dynamic evolution is 300 time steps. }
\vspace{-0.3cm}
\label{fig:Fig_30}
\end{figure}

One of the most pivotal human characteristics for enhanced community resilience is flexibility. Figure ~\ref{fig:Fig_30} displays dynamic changes of fear, information-seeking behavior, and risk perception as a result of the changes in individuals$\prime$ flexibility. Flexibility has a direct effect on emotion and risk perception, while it has an indirect impact on information-seeking behavior. It is obvious that when the flexibility increases (from $(M^{F}=0)$ to $(M^{F}=0.5)$ to $(M^{F}=1)$), individuals demonstrate a lower level of fear. More flexible people are able to more thoroughly evaluate their emotions so that, consequently, the level of fear and depression is decreased \cite{thayer2009heart}. Negative affects\footnote{In social science, emotion and affect are considered to be similar words to each agent's response to feelings \cite{Barsade1998}.}, like the feeling of fear, make a person less flexible in interpersonal cognition and expressive behavior. Conflict is caused in discussions among startled people. In other words, flexibility is diminished among these individuals \cite{hollenstein2006state}. In contrast, a high level of flexibility and a low level of fear decrease the perceived risk of agents during a disaster. As a result, information-seeking behavior which is profoundly entwined with risk perception is reduced.\par
Conversely, the feeling of fear causes people to be flexible if they are optimistic. That is why flexibility is increased at the beginning of the event when $(M^{F}=0)$. Because all of the individuals of the community mentioned above are optimistic, they tend to be more flexible during the first time interval. In general, positive features can disguise a person's behavioral drawbacks. Since the news from the mass media is often related and stressful, the average emotional changes increase over time, no matter how much flexibility there is. Correspondingly, the level of risk perception and information-seeking behavior of agents will increase.

\vspace{-0.3cm}
\subsection{Effects of Cooperation on Human Responses}

\begin{figure}
\centering
\includegraphics[width=1 \columnwidth]{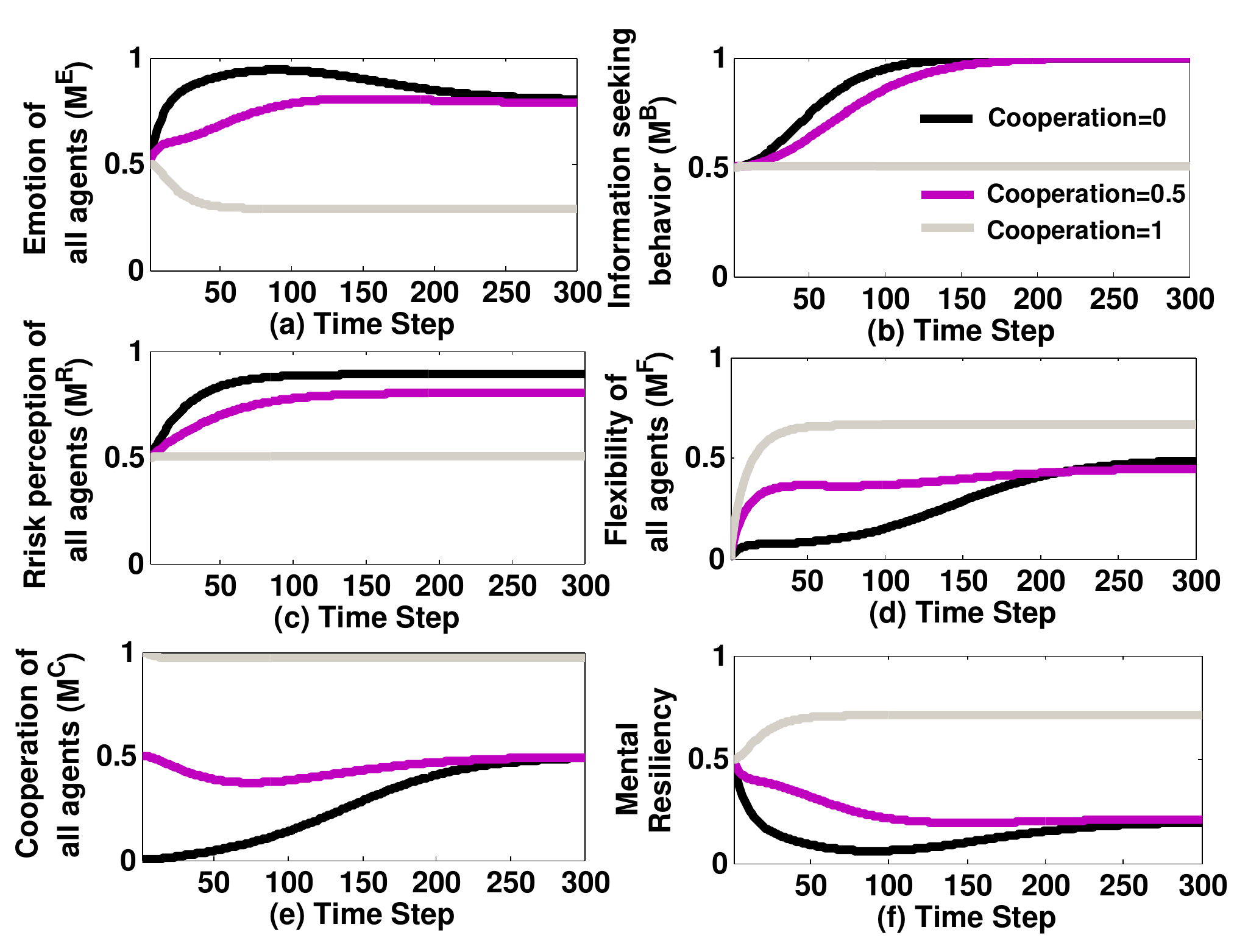}
\setlength{\abovecaptionskip}{-15pt}
\caption{Effects of the different initial values of cooperation on the dynamic change of emotion (panic in this study), information-seeking behavior, risk perception, flexibility, and mental resilience (mental well-being). The initial value of cooperation is assumed to be 0, 0.5, and 1. These figures are related to the collective behavior of all agents. }
\vspace{-0.6cm}
\label{fig:Fig_31}
\end{figure}

Another human aspect that is vital during a disaster is cooperation. Cooperation is somewhat similar to flexibility in that they both have a direct influence on fear as well as risk perception. Furthermore, cooperation affects flexibility and vice versa. Figure ~\ref{fig:Fig_31} shows changes in fear, information-seeking behavior, risk perception, and flexibility induced by changes in the cooperation of individuals. When the level of cooperation increases from $M^{C}=0$ to $M^{C}=0.5$  to $M^{C}=1$, the agents experience a lower level of fear, risk perception, and information-seeking behavior. Flexibility, in contrast, increases. High levels of cooperation can make positive changes in individual behavior \cite{fowler2010cooperative}.\par
Since the level of fear is less than the fear threshold when $M^{C}=1$ and the level of cooperation is high in the ideal case, risk perception shows no noticeable increase during a crisis. The level of information-seeking behavior roughly follows the same trend as the risk perception. When $M^{C}=0$, the perceived risk, the  level of fear, and the information-seeking behavior of agents increases. \par
On the other hand, the feeling of fear during disasters makes humans  cooperate and be more flexible. Hence, when the individual cooperation and flexibility of individuals are augmented, the level of fear deceases. However, a decrease in the level of fear results in a lower level of cooperation. Of note, the loop of fear, cooperation, and flexibility formed here is directly or indirectly impacted by the mass media, the accessibility of electricity and emergency services and other mental peculiarities during dynamic change. According to Figure ~\ref{fig:Fig_31}(b), an increase in cooperation among individuals in the community is associated with an increment in mental resilience or well-being. Thus, the society with more mental well-being has more community resilience.

\vspace{-0.3cm}
\subsection{Effects of Experience on Human Responses}

\begin{figure}
\centering
\includegraphics[width=1 \columnwidth]{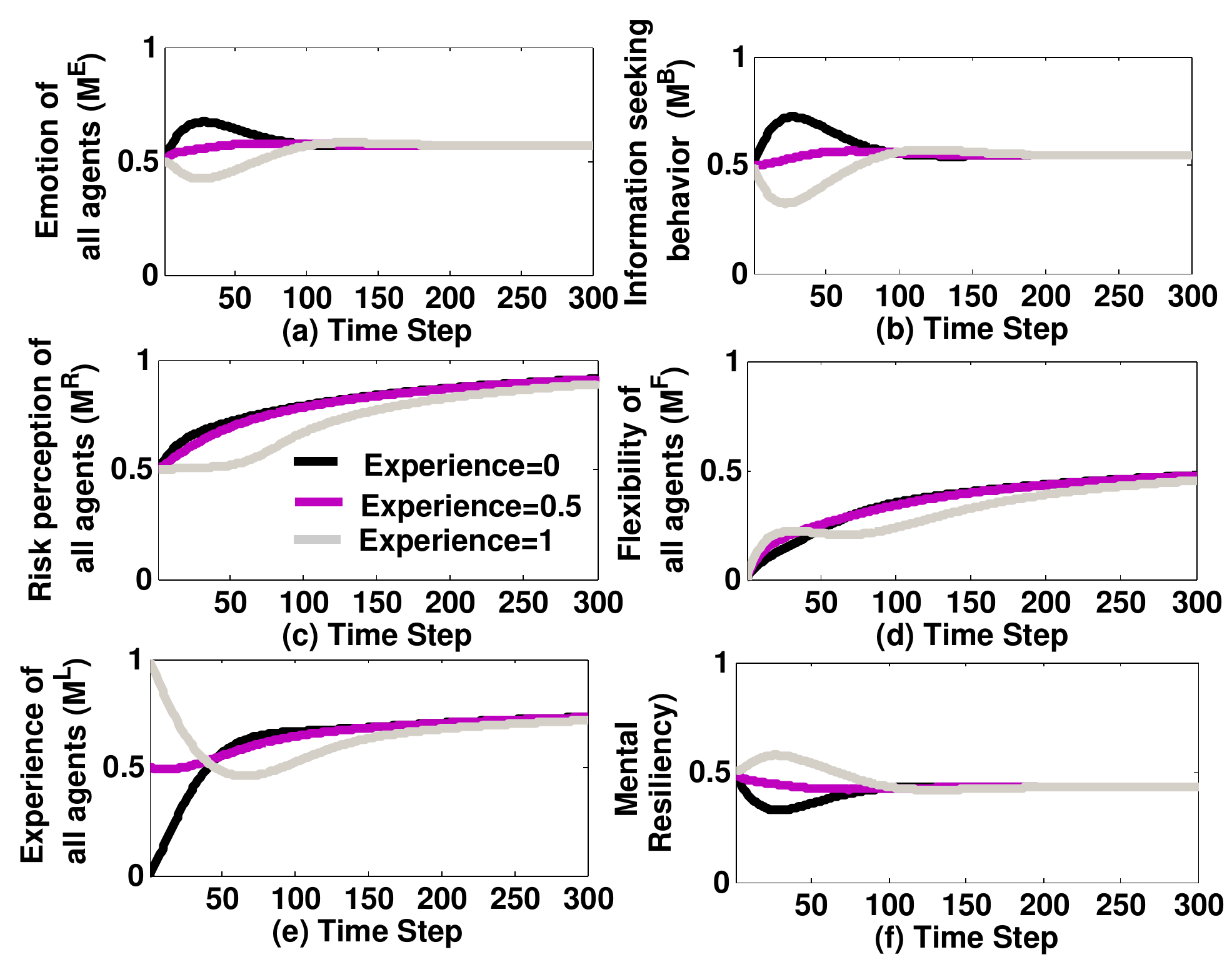}
\setlength{\abovecaptionskip}{-15pt}
\caption{Effects of experience on other human characteristics in a community with similar behavior. To carry out the sensitivity analysis, the initial value of experience is assumed to be equal to 0, 0.5, and 1. In this figure, the dynamic change of emotion, risk perception, information-seeking behavior, flexibility, experience, and mental resilience of all agents involved in the community is presented for the time interval [0 300].}
\vspace{-0.6cm}
\label{fig:Fig_32}
\end{figure}

People with previous experience in disasters can cope with the fear from a disaster easier than the ones without experience. Additionally, people who have experienced special disasters have a higher level of risk perception with respect to this event happening. Figure ~\ref{fig:Fig_32} shows the changes in fear, information-seeking behavior, risk perception, and flexibility in tandem with changes in the experience levels of individuals. Experience has an inverse effect on fear, information-seeking behavior, and risk perception. It positively affects flexibility if agents are optimistic.\par
Because all people of this community are optimistic, the feeling of fear of agents tends to reach the same level when the initial level of experience is different. When the agents do not have previous experience, they seek new information during a disaster. Therefore, their experience increases. If the initial experience is low, fear and risk perception increase in the first time interval, while flexibility reduces compared to when the initial experience is higher. \cite{ hormiga2014relationship } discusses the correlation between experience and risk perception. The society that has previous experience with one particular hazard perceives a higher level of risk than the society without previous disasters. There is a stable feedback mechanism between experience and risk perception \cite{ ohman2017previous} in the cognitive mechanism. Furthermore, flexibility and experience can lead to distinguished achievements \cite{gelb2008knowing}. If people have more experience at the beginning of the disaster, they have more mental resilience and social well-being. Because people gain experience during disasters, society becomes more resilient compared to when the disaster first occurred. Of note, experience of agents with the uncertainty based-opinion is increased faster than informed agents \cite{cho2018dynamics}.

\vspace{-0.3cm}
\subsection{Effects of Cooperation and Experience on Human Responses}

\begin{figure}
\centering
\includegraphics[width=1 \columnwidth]{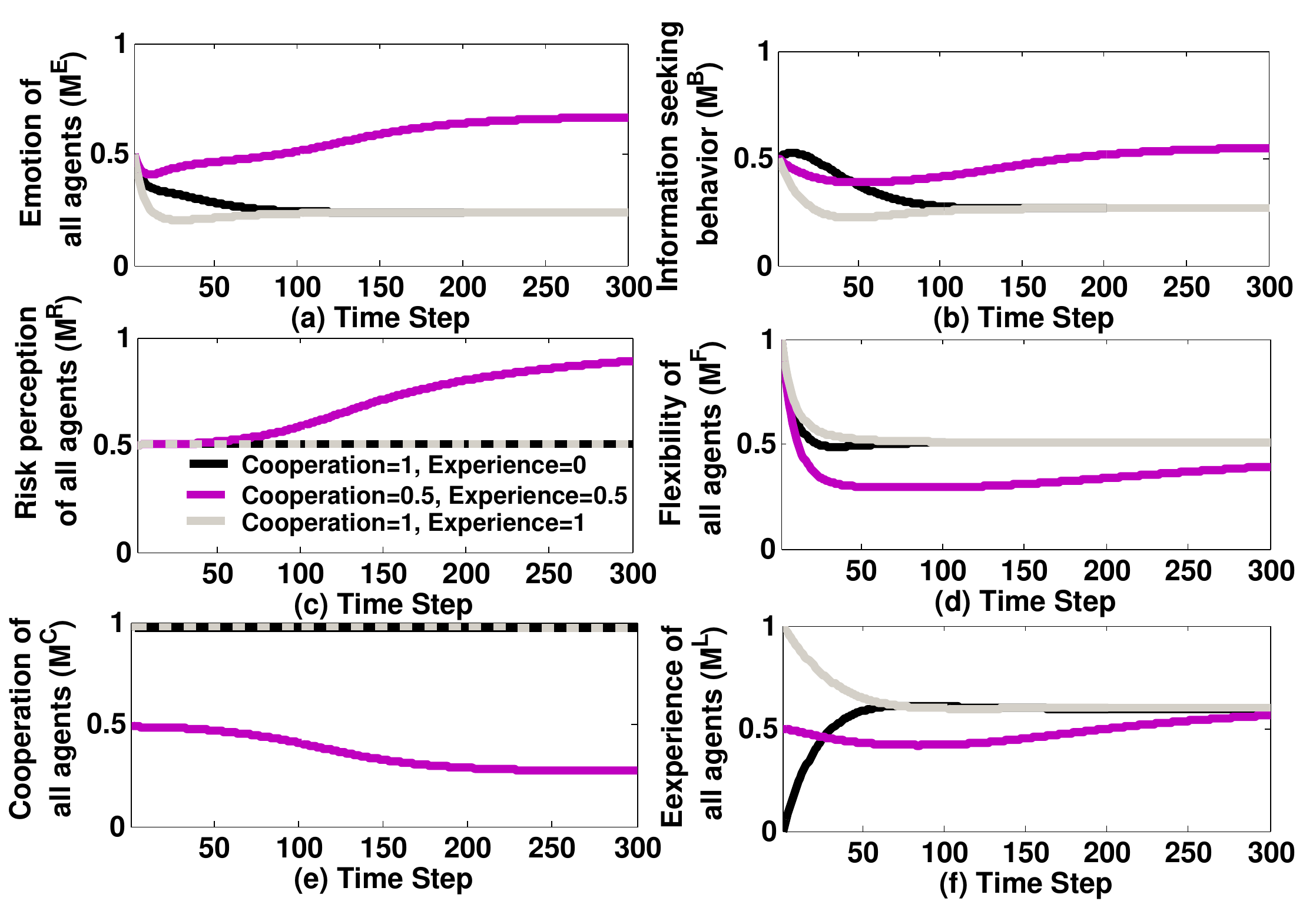}
\setlength{\abovecaptionskip}{-15pt}
\caption{Effects of the different initial values of cooperation and experience on the dynamic change of the collective mental behavior in the homogeneous community. The black lines denote when people are well-experienced and enthusiastic to cooperate. On the other hand, grey denotes agents that are not interested in cooperating at all. The purple lines represent individuals, who are only partially experienced and for whom the level of enthusiasm to cooperate is not high.    
}
\vspace{-0.57cm}
\label{fig:Fig_33}
\end{figure}

Figure ~\ref{fig:Fig_33} presents changes in fear, information-seeking behavior, risk perception, and flexibility with respect to changes in the cooperation and experience of individuals. Three different examples are provided. In Example 1, although people are willing to cooperate, they do not have previous experience regarding the disaster. In Example 2, both $M^{L}$ and $M^{C}$ are equal to 0.5 while they are  equal to 1 in Example 3. \par
In Examples 1 and 3, when $M^{L}$ =1, a high level of cooperation and optimism lead to a low level of fear such that panic is lower than the fear threshold. In Example 3, since the agents have a high level of cooperation and experience, they do not feel a need to seek new information. Additionally, individuals are more flexible than the individuals in examples 1 and 2. In examples 1 and 3, because of the low level of fear, the level of risk perception and cooperation among agents do not show substantial variations. The level of experience of the agents in Example 1 is higher than that in Example 2, resulting from higher levels of cooperation among individuals. Risk perception and individuals' information-seeking behavior hinge upon cooperation \cite{ring1992structuring}.  In perilous circumstances, agencies raise public risk perception to levels that exceed what individuals experience privately. According to \cite{ring1992structuring}, the obstacles to private-private cooperation are more than those that individuals experience with  private-public cooperation. 

\vspace{-0.3cm}
\subsection{Effects of Cooperation on Electric Energy Sharing}

\begin{figure}
\centering
\includegraphics[width=1 \columnwidth]{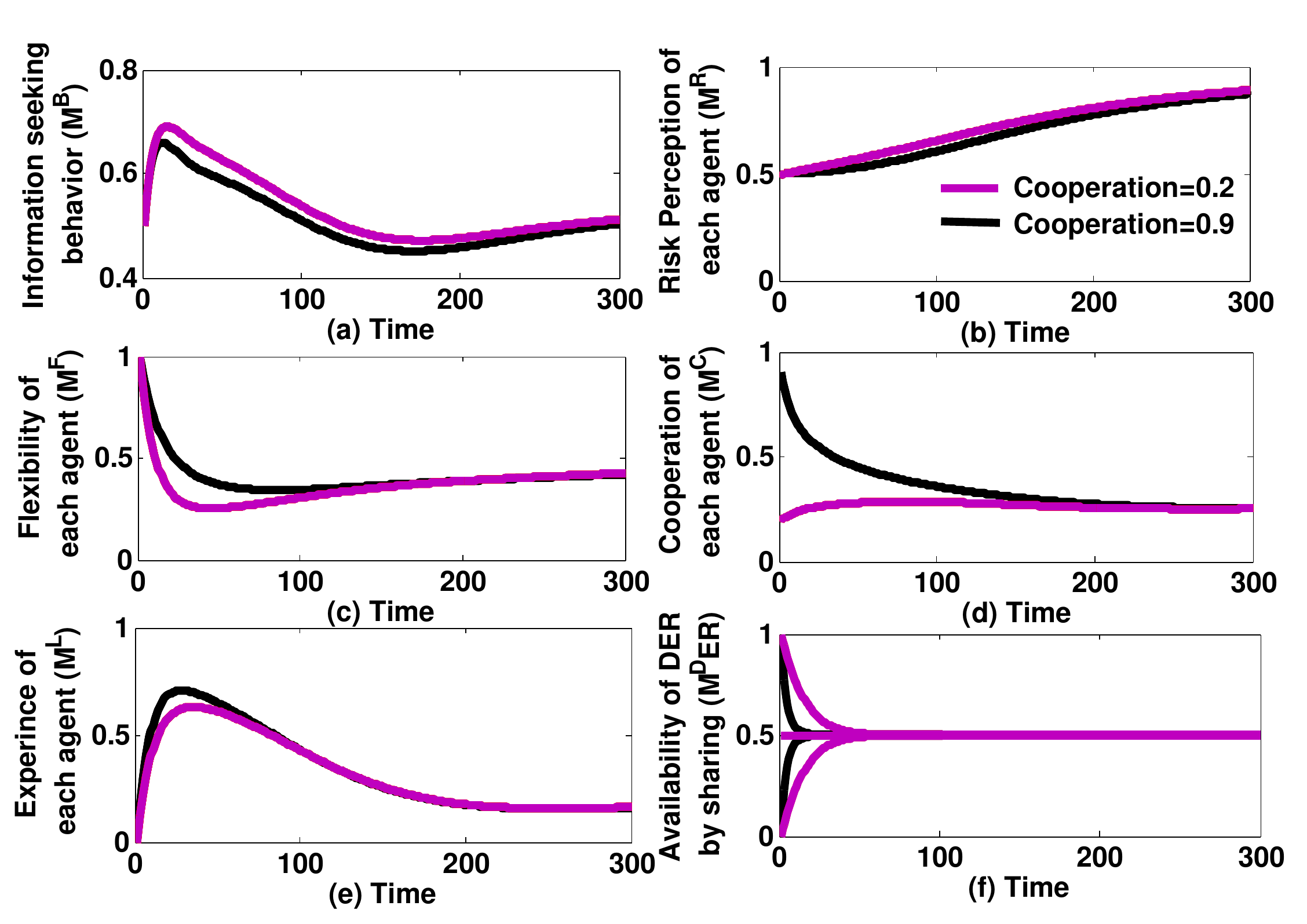}
\setlength{\abovecaptionskip}{-15pt}
\caption{Effects of cooperation on electricity sharing and the impact of the availability of electricity (and also cooperation) on information-seeking behavior, risk perception, flexibility, and experience. It is assumed that agents have a varying value of accessibility to distributed energy resources, i.e., 0, 0.5, and 1. The results are provided for initial values of cooperation of 0.2 (low cooperation) and 0.9 (high cooperation).  }
\vspace{-0.6cm}
\label{fig:Fig_35}
\end{figure}

\begin{figure}
\centering
\includegraphics[width=1 \columnwidth]{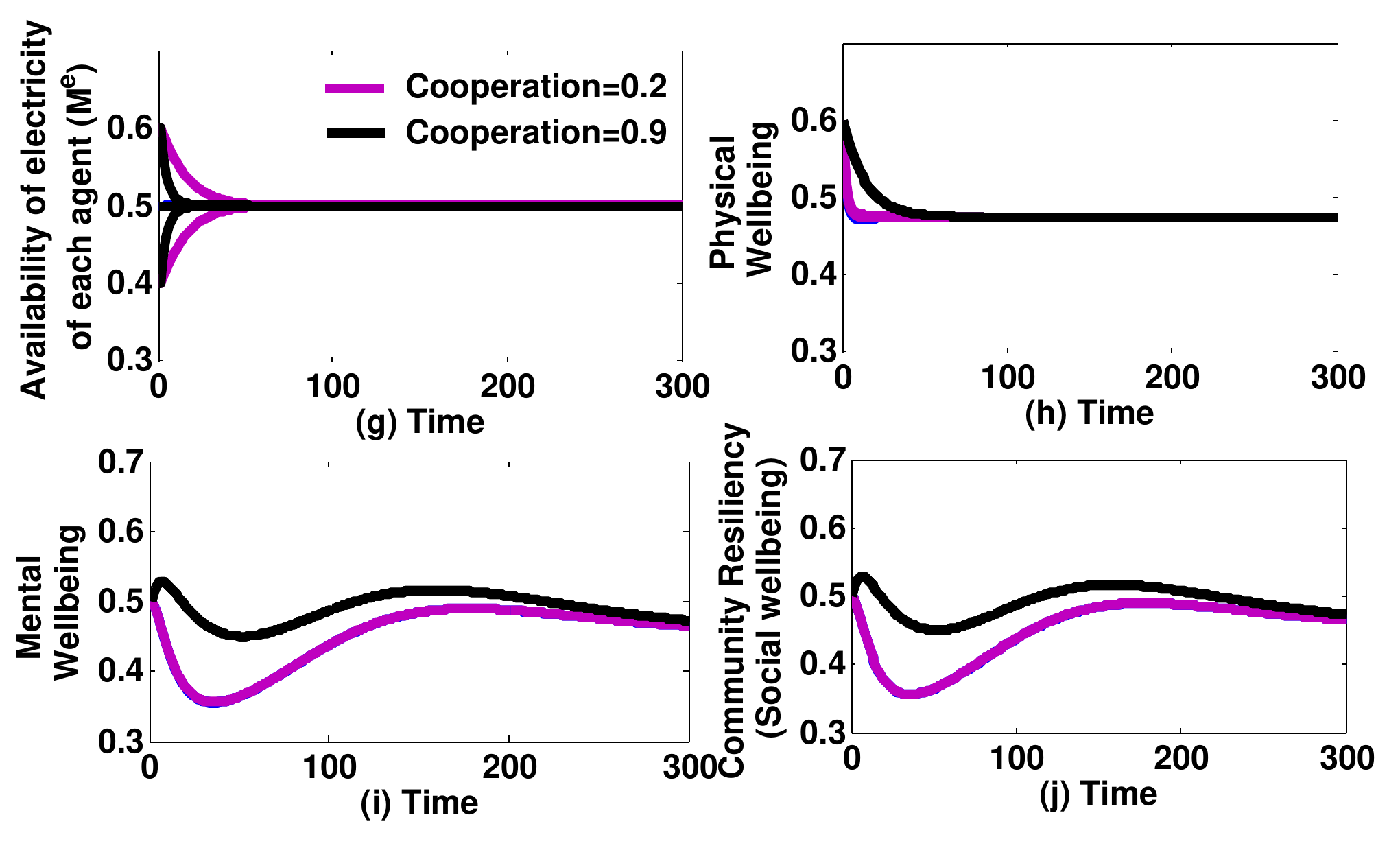}
\setlength{\abovecaptionskip}{-15pt}
\caption{Effects of different initial values of cooperation on the availability of electricity, physical well-being, mental well-being, and community resilience. Results are provided for different levels of cooperation (0.2 and 0.9). In this homogeneous community, the accessibility of agents to DERs varies. The dynamic change of all kinds of well-being is provided for the time interval [0 300].
}
\vspace{-0.6cm}
\label{fig:Fig_35b}
\end{figure}

To investigate the effect of the level of cooperation on electricity sharing, the availability of electricity from distributed energy resources for three agents within each area are assumed to be 0, 0.5, and 1, respectively. The results are provided for two different levels of cooperation in Figures ~\ref{fig:Fig_35} and ~\ref{fig:Fig_35b}. According to these results, when people have a high level of cooperation, they share their electricity sooner than when they have a low level of cooperation. Consequently, they have a higher level of physical health when $M^{C}=0.9$. Furthermore, with a high level of cooperation and physical health, people experience less panic. As a result, the level of perceived risk and information-seeking behavior among agents is decreased compared to  when $M^{C}=0.2$. However, the level of panic climbs with time as a result of relevant negative news from the mass media. Thus, when $M^{C}=0.9$, the level of flexibility drops after its initial growth. These factors make the average cooperation lower over time. In addition, a society with more cooperation has a higher level of physical and mental well-being and community resilience (social well-being) when there are both prosumers and consumers in the community. Note that both cooperative and selfish behavior among individuals are assumed to be epidemic \cite{che}. Furthermore, according to \cite{chen2017fundamental}, cooperation is of high importance for a successful society in both fixed (static) social networks and fluid (dynamic) social network. The social diffusion of cooperation exists in both kinds of networks.

\vspace{-0.3cm}
\subsection{Importance of Emergency Services, the Injury Factor of a Disaster, and News Polarity on Physical and Emotional Well-Being}
\begin{figure}
\centering
\includegraphics[width=1 \columnwidth]{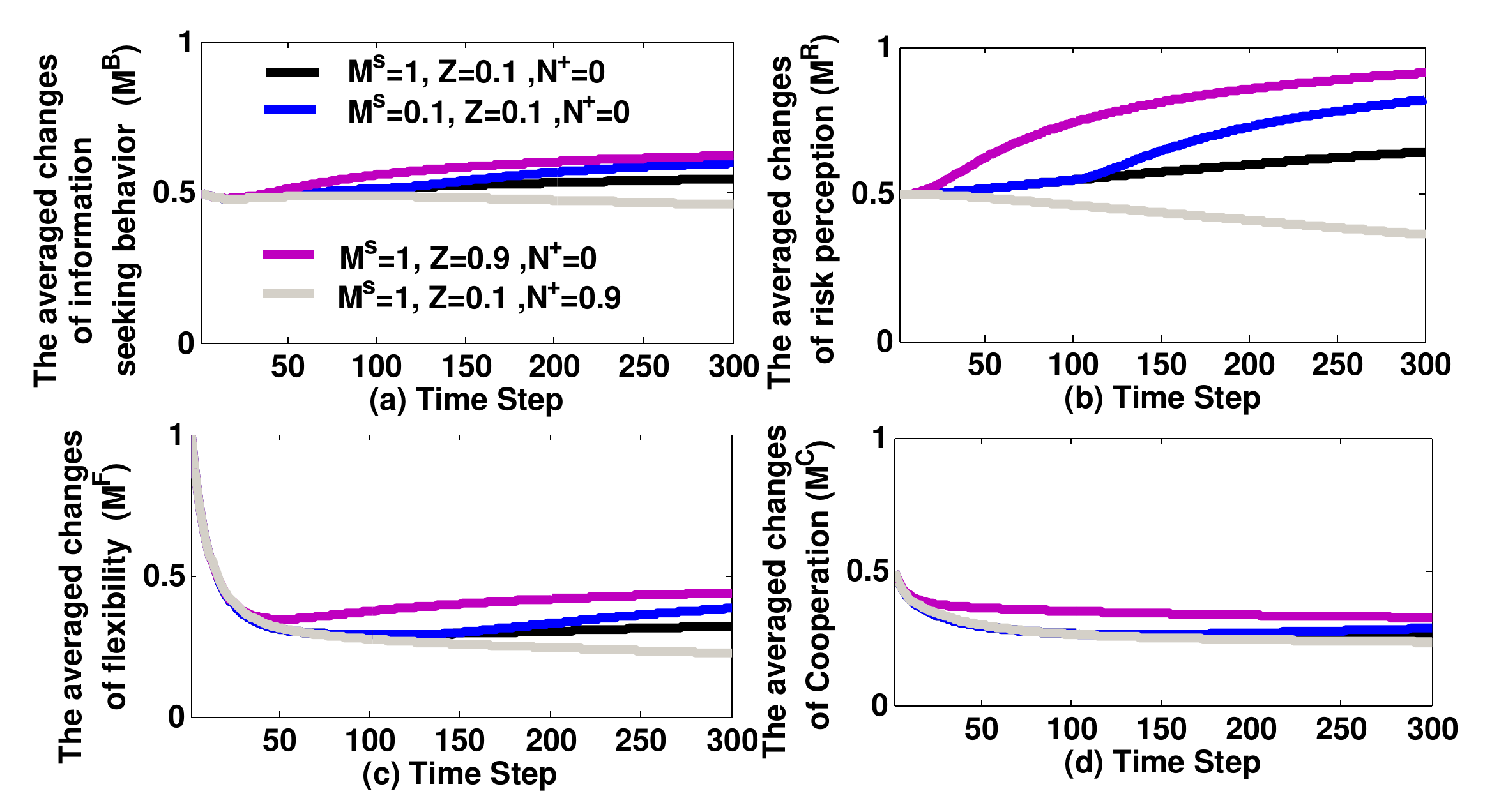}
\setlength{\abovecaptionskip}{-15pt}
\caption{Effects of emergency services, the injury factor of disaster, and news polarity on collective information-seeking behavior, risk perception, flexibility, and cooperation. The black lines represent the effects when emergency services are entirely available, and the disaster is benign. However, there are a lot of rumors and negative news among individuals. In contrast, the  blue lines represent the effects when emergency services are not available to the community. In the case of the purple lines, emergency services are wholly available but the disaster is severe, and there is positive news at the level of the society. The grey lines also represent a case when emergency services are available. However, in this case, there is lots of positive news and the disaster is not severe}.  
\vspace{-0.6cm}
\label{fig:Fig_36}
\end{figure}

\begin{figure}
\centering
\includegraphics[width=1 \columnwidth]{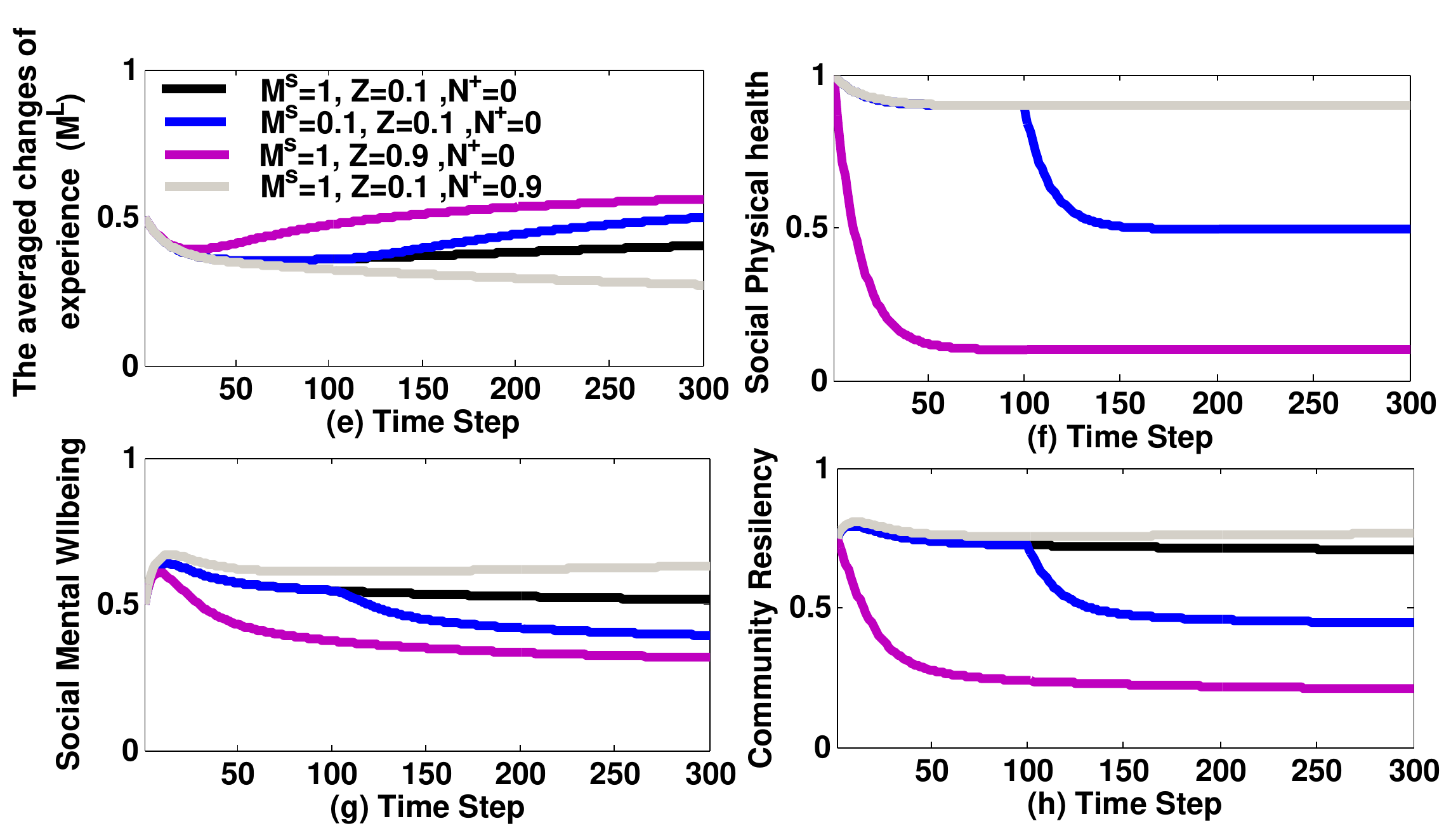}
\setlength{\abovecaptionskip}{-15pt}
\caption{Effects of emergency services, the injury factor of disaster, and news polarity upon experience, physical and mental well-being, and community resilience. The grey lines represent the effects when all outside factors, i.e., emergency services, the disaster, and the mass media, do not have a negative effect on the community. Understandably, there is more community resilience in this case compared to that of other scenarios for time interval [0 300].  }
\vspace{-0.6cm}
\label{fig:Fig_36b}
\end{figure}

In this section, the effect of emergency services, the injury factor of a disaster, and news polarity on collective mental features, physical and emotional well-being, and community resilience is investigated. Results are shown in Figures ~\ref{fig:Fig_36} and ~\ref{fig:Fig_36b}. In Example 1, $M^{S}=1$, Z=0.1 and $N^{+}=0$.
 $M^{S}$ is assumed to be 0.1 from time stamp 100 to 300 in Example 2. To show the effect of the injury factor of disaster, Z is 0.9 in Example 3. To present the effect of news polarity, $N^{+}$ is 0.9 in Example 4. \par
In Example 2, because of the disaster, a lack of appropriate emergency infrastructure, the destruction of part of the emergency facilities during an event, and a shortage of emergency staff since time stamp 100, emergency services cannot effectively perform their function. When emergency services decrease, the average physical health of individuals sharply declines. Therefore, the agents' level of fear increases from time step 100 on, and, in turn, risk perception and information-seeking behavior increases. As a consequence, individuals obtain more experience and become more flexible compared to when emergency services are sufficiently available.  \par
In Example 3, when the injury factor of disaster is very high with a value of 0.9, the physical health of individuals is dramatically lower. This case is similar to Example 2 in that the trends of fear, information-seeking behavior, and risk perception are the similar. Note that the human response in case 3 changes more quickly than in Example 2. In addition, due to the high level of community fear, the level of cooperation among individuals grows. \par
In Example 4, the news from the mass media is positive with a value of 0.9, and the level of fear, perceived risk, and information-seeking drops. For this example, the community obtains less experience as compared to Examples 1, 2, and 3. In addition, individuals tend to be less flexible compared to when they feel endangered. \par

\vspace{-0.3cm}
\subsection{Effects of Different Mass Media Trends on Community Resilience}

\begin{figure}
\centering
\includegraphics[width=1 \columnwidth]{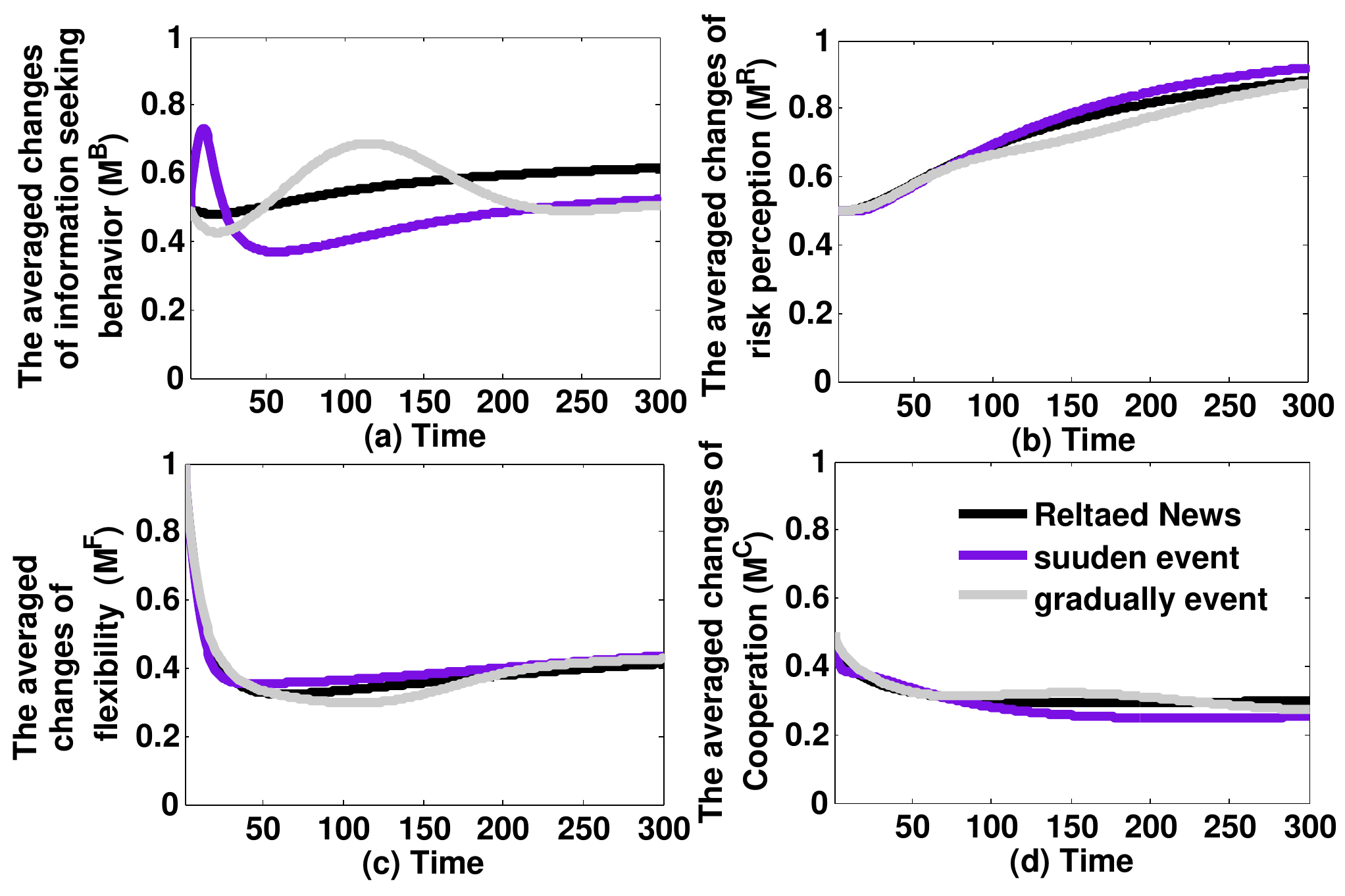}
\setlength{\abovecaptionskip}{-15pt}
\caption{Effects of different mass media trends on collective information-seeking behavior, risk perception, flexibility, and cooperation. The black lines are associated with all the news being sensational during the entire interval [0 300]. For the purple lines, the mass media trends follow a damped exponential model (sudden events), while for the grey lines the mass media trends follow an exponential model (gradual events).}
\vspace{-0.6cm}
\label{fig:Fig_37}
\end{figure}

\begin{figure}
\centering
\includegraphics[width=1 \columnwidth]{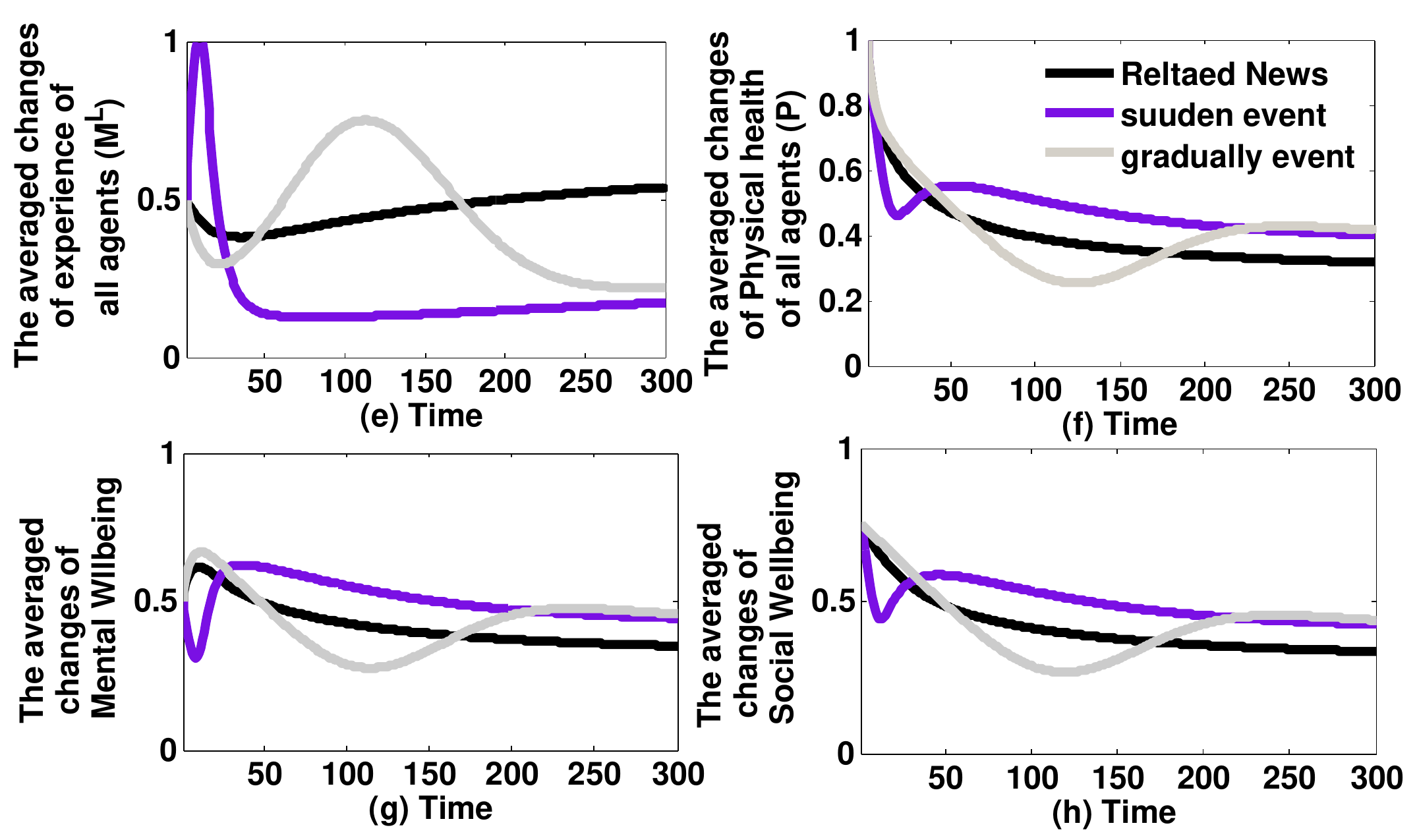}
\setlength{\abovecaptionskip}{-15pt}
\caption{Effects of different mass media trends on the dynamic change of collective experience, physical health, mental well-being, and social well-being (community resilience) for the time interval [0 300]. Various mass media trends influence community resilience differently. As a result, the importance of the mass media on community resilience can be easily grasped.  }
\vspace{-0.6cm}
\label{fig:Fig_37b}
\end{figure}

Figures ~\ref{fig:Fig_37} and ~\ref{fig:Fig_37b} show the dynamic changes in the mental characteristics and physical health for different media values. In Example 5, all news is assumed to be related. Examples 6 and 7 are related to sudden events (tsunamis, earthquakes) and gradually unfolding events (like a hurricane and social crisis), respectively. The fitting function for the mass media are $N_{t}=2.5*e^{(-3*t)}+0.04$,
$N_{t}=1*e^{-(\frac{(t-50)^{2}}{50})}+0.06$
, respectively. For events which happen suddenly, the feeling of panic is stimulated at the beginning of the disaster. As a consequence, individuals seek more information during this time. Understandably, they also obtain more experience. On the other hand, when the event occurs gradually, the level of fear, information-seeking behavior, and risk perception of the community is raised during the midterm.  


\vspace{-0.3cm}
\subsection{Effects of Compassionate Empathy on Collective Behavior During and After a Disaster}

\begin{figure}
\centering
\includegraphics[width=1 \columnwidth]{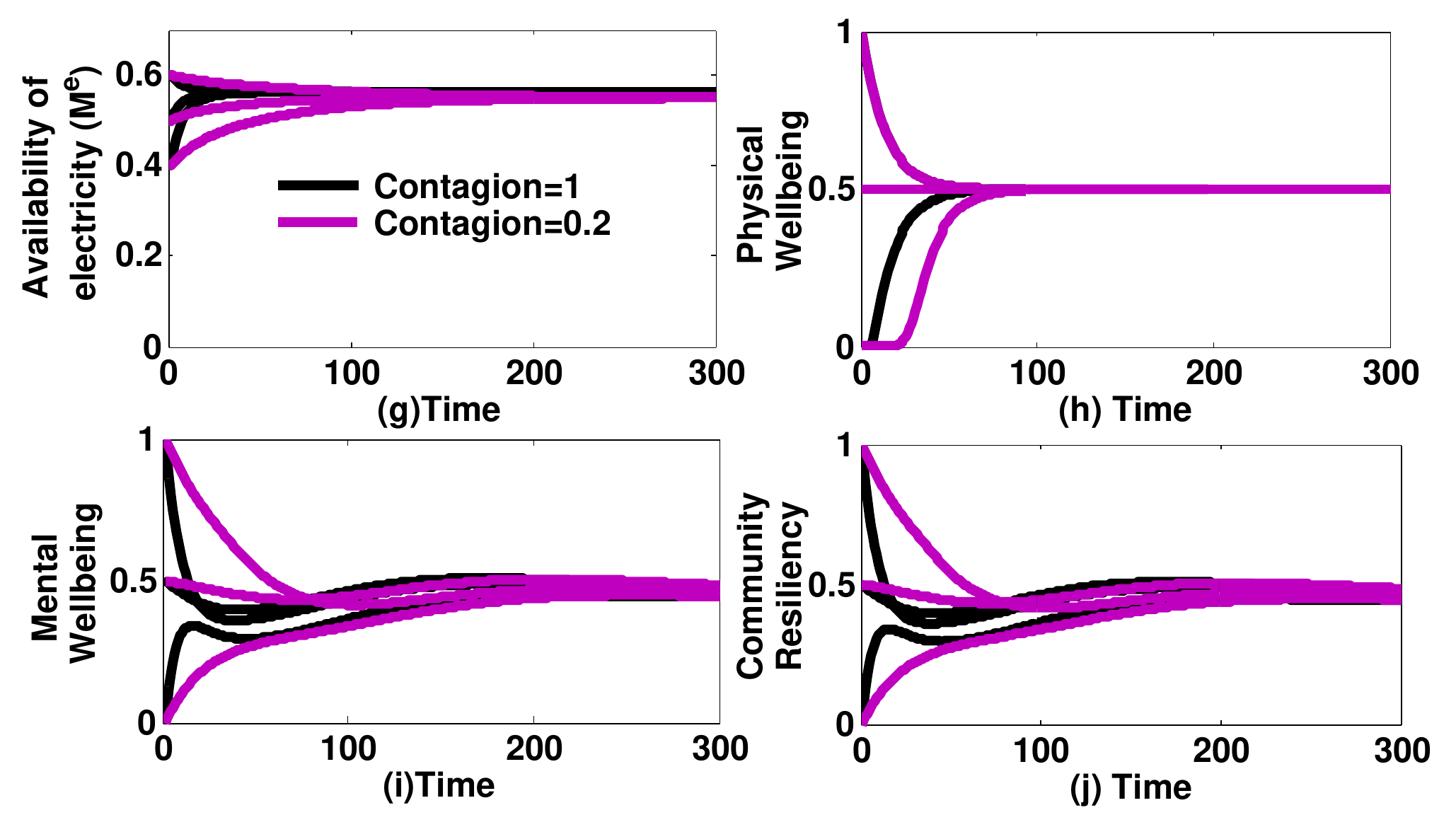}
\setlength{\abovecaptionskip}{-15pt}
\caption{Effects of compassionate empathy on the level of availability of electricity in the community, physical well-being, mental well-being, and community resilience (social well-being). The difference between outputs of the two examples is related to the time they begin to assist each other.   }
\vspace{-0.6cm}
\label{fig:Fig_38b}
\end{figure}

Figures ~\ref{fig:Fig_38} and ~\ref{fig:Fig_38b} show the dynamic changes in human response and community resilience for a different levels of compassionate empathy among individuals. To clarify the importance of empathy on community resilience, the human mental and physical characteristic of 3 agents inside each area are assumed to be 0\%, 0.5\%, and 1\%, respectively. The results are presented for two different levels of empathy, 0.1 and 1. Less empathy among individuals plus other characteristics, including fear, information-seeking behavior, flexibility, and cooperation, all tend to converge at the same level. Also, agents share their electricity later than when compassionate empathy is 1. As a result, when the empathy is high, the average level of physical well-being, mental well-being, and community resilience for the whole time interval [0 300] is more than that when individuals are not empathetic.

\begin{figure}
\centering
\includegraphics[width=1 \columnwidth]{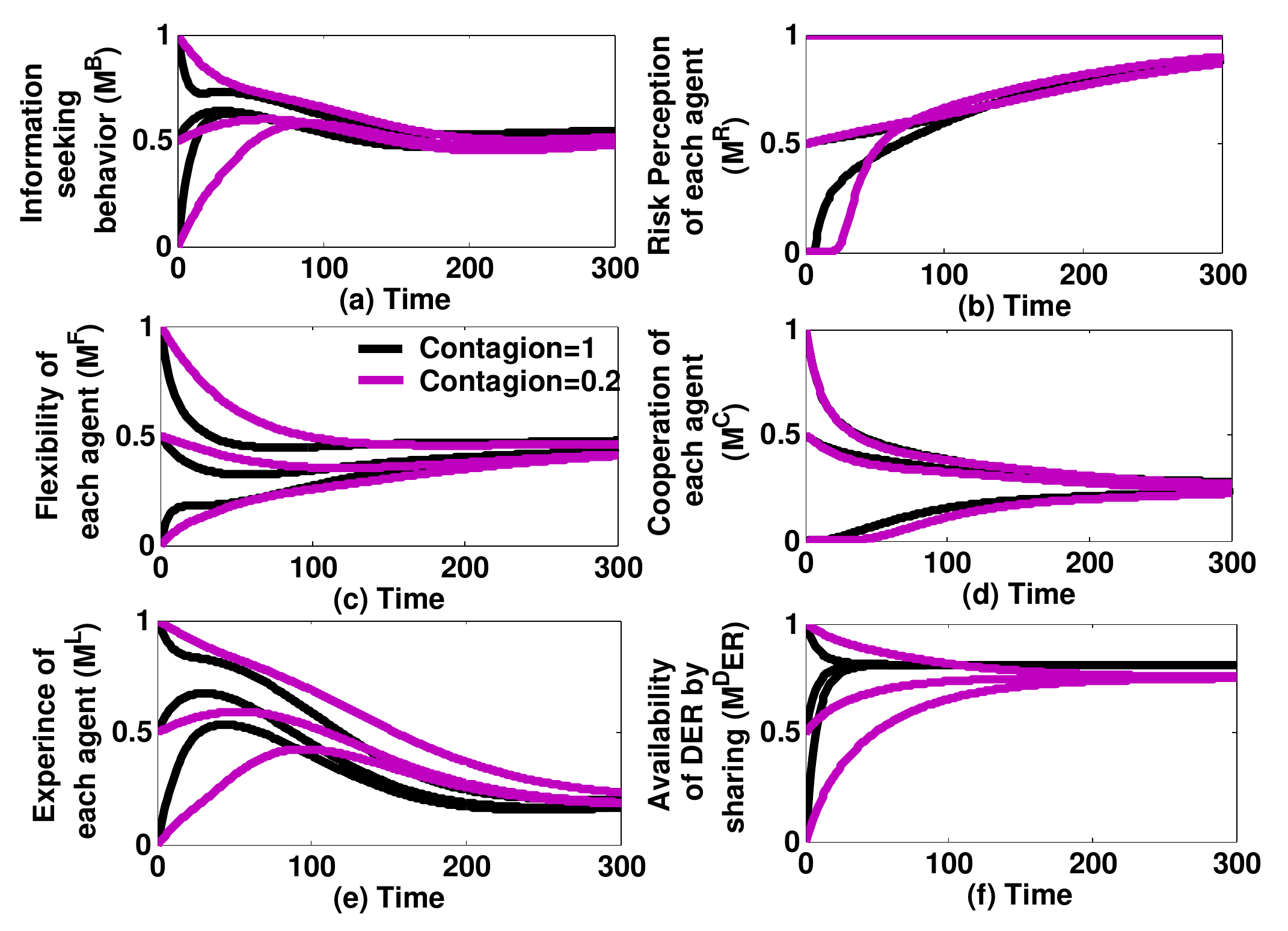}
\setlength{\abovecaptionskip}{-15pt}
\caption{Effects of different values of compassionate empathy on collective information-seeking behavior, risk perception, flexibility, cooperation, experience, and availability of electricity supplied by DERs. Two different values of compassionate empathy are 0.2 (low level of empathy among individuals) and 1 (high level of empathy among individuals).  }
\vspace{-0.6cm}
\label{fig:Fig_38}
\end{figure}

\vspace{-0.2cm}
\section{Simulation Results for Case Study 2: Society of Six Separate Communities}
This case study aims to clarify the effect of the different scenarios of disasters (in terms of time, place, and kind of emergency) on the mental well-being, physical well-being, and community resilience. Three different scenarios in this case study are analyzed. These scenarios are presented in Table~\ref{tab:FC101}. In addition, the social effect of the diversified community population on its social well-being during and after a disaster is analyzed.\par
Figure ~\ref{fig:Fig_38} depicts a society, consisting of six communities with different characteristics. The parameter setting for the mental and physical characteristics, population, and electric grid related to each Community are provided in Table~\ref{tab:FC10}. The level of intra- and inter-community empathy is shown in Table~\ref{tab:FC100}. It is found that Communities 1 and 2 are extremely close-knit. As a result, empathy among these communities are assumed to follow the Gaussian distribution $N(0.9,0.1^{2})$. Regarding the other communities, its is assumed that there is no empathy among them. \par 

\begin{table}[H]
\caption{ Three scenarios of disasters in terms of time, place, and type of emergency.}
\centering
\scriptsize
\begin{tabular}{|p{0.7cm}|p{7.25 cm}|} 
\hline  
Scenario & Emergency feature \\ \hline
1&  The occurrence of one disaster in a community \\ \hline
2 & The occurrence of two concurrent disasters in two different communities\\ \hline
3 & Emergencies that arise at different times\\ \hline
\end{tabular}
\label{tab:FC101}
\vspace{-0.6cm}
\end{table}

\begin{table}[H]
\caption{Parameter settings for the community characteristic of the second case study i.e., the society of six separate communities, where $C_{i}$ means community $i (i \in 1,2,3,4,5,6$).}
\resizebox{0.9\textwidth}{!}{\begin{minipage}{\textwidth}
\scriptsize
\begin{tabular}{|p{1 cm}|p{1.6 cm}| p{1.6 cm}| p{3.5 cm}|}
\hline  
Parameter & $C_{1}$ & $C_{2}$ & $C_{3}$, $C_{4}$, $C_{5}$ and $C_{6}$  \\ \hline
$M^{R}_{ti}$& $N(0.8,0.1^{2})$ & $N(0.7,0.1^{2})$ &  $N(0.1,0.1^{2})$ \\ \hline 
$M^{B}_{ti}$& $N(0.8,0.1^{2})$ & $N(0.7,0.1^{2})$  &  $N(0.1,0.1^{2})$ \\ \hline
$M^{E}_{ti}$ & $N(0.98,0.02^{2})$ & $N(0.1,0.1^{2})$ &  $N(0.1,0.1^{2})$ \\ \hline
$M^{F}_{ti}$ & $N(0.5,0.1^{2})$ &$N(0.5,0.1^{2})$&$N(0.5,0.1^{2})$ \\ \hline
$M^{L}_{ti}$ & $N(0.5,0.1^{2})$ &$N(0.5,0.1^{2})$ &$N(0.5,0.1^{2})$ \\ \hline
$M^{C}_{ti}$ & $N(0.5,0.1^{2})$ & $N(0.5,0.1^{2})$ & $N(0.5,0.1^{2})$ \\ \hline
$P_{ti}$ & $N(0.5,0.1^{2})$ & $N(0.98,0.02^{2})$& $N(0.98,0.02^{2})$\\ \hline
Population& 150 & 250& 135, 450, 500, and 120\\ \hline
\end{tabular}
\label{tab:FC10}
\end{minipage}}
\vspace{-0.6cm}
\end{table}

\begin{table}[H]
\caption{Levels of intra- and inter-communities empathy. }
\centering
\scriptsize
\begin{tabular}{|p{1.5 cm}|p{1.5 cm}| p{1.5 cm}|p{1.5 cm}|}
\hline  
Community & $C_{1}$ & $C_{2}$ & $C_{i(i\in3,4,5,6)}$ \\ \hline
$C_{1}$&$N(0.9,0.1^{2})$&$N(0.9,0.1^{2})$&-  \\ \hline
$C_{2}$ &$N(0.9,0.1^{2})$&$N(0.9,0.1^{2})$&- \\ \hline
$C_{i(i\in3,4,5,6)}$ &-&-&$N(0.9,0.1^{2})$\\ \hline
\end{tabular}
\label{tab:FC100}
\vspace{-0.6cm}
\end{table}

\begin{figure}
\centering
\includegraphics[width=1\columnwidth]{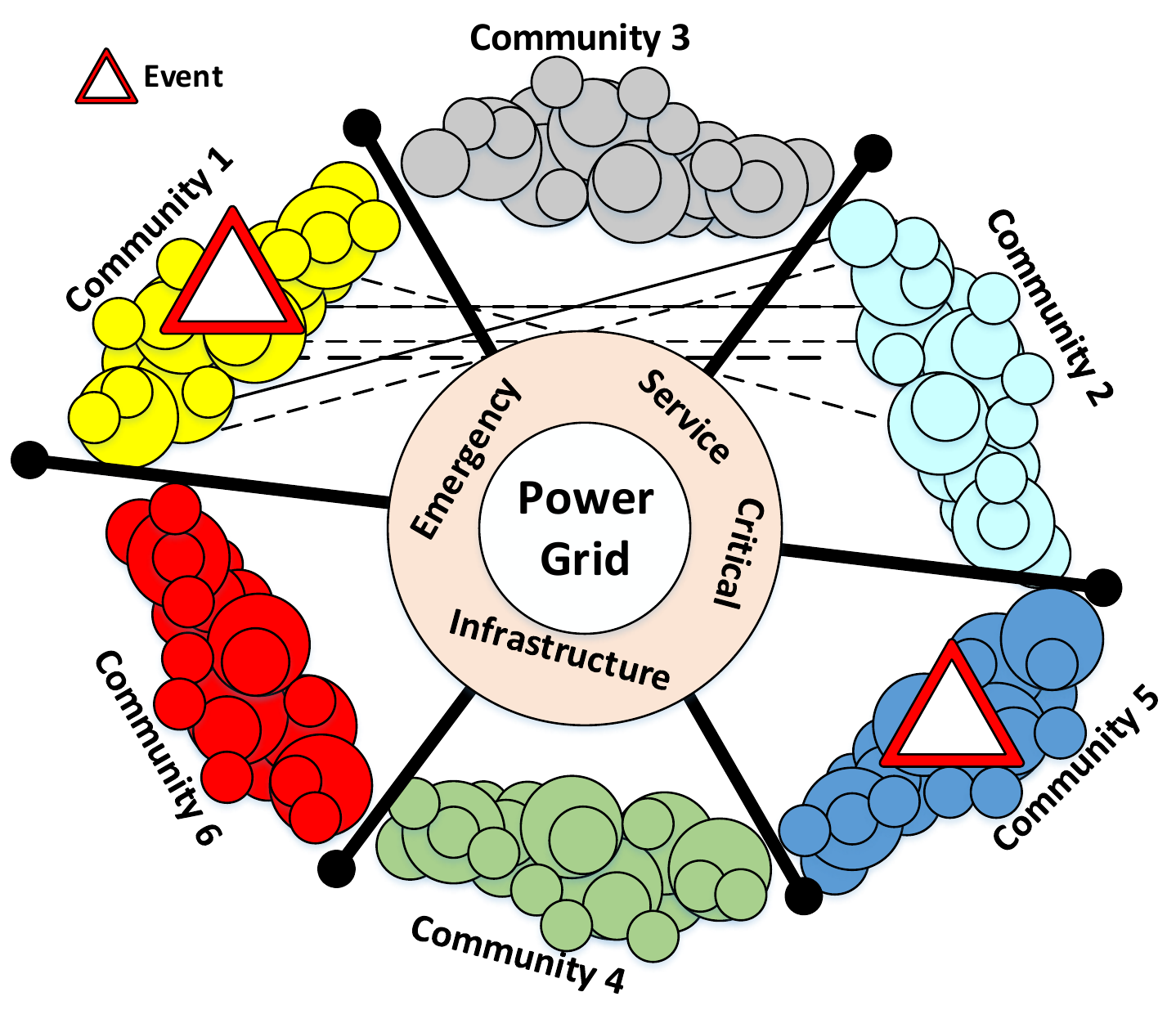}
\setlength{\abovecaptionskip}{-15pt}
\caption{A society consisting of six separate communities. It is assumed that individuals in Communities 1 and 2 are empathetic to each other. Two critical infrastructures, i.e., power system and emergency services, are considered to support communities during disasters. The availability of each of these two critical infrastructures in each community highly depends on the kind of disaster that occurred in the community. Additionally, there are two potential places the where the disaster occurs directly (i.e., in Communities 1 and 5).}
\vspace{-0.6cm}
\label{fig:Fig_38}
\end{figure}

\vspace{-0.3cm}
\subsection{Effects of the Occurrence of a Disaster on Human Response}
Each disaster can be modeled with the distinct characteristics of Z, $Q^{s}$, $Q^{e}$, and $M^{e}$. In Example 1, the disaster only occurs in community 1. The injury factor of disaster is assumed to be $N(0.9,0.1^{2})$. Because of severe hazards, emergency services and the power utility are inaccessible in community 1, but the individuals in this community can still utilize on-site generation. $Q^{DER}_{ti}$  follows the Gaussian distribution $N(0.5,0.1^{2})$. In other communities, $Q^{s}_{ti}$, $Q^{e}_{ti}$, and $Q^{DER}_{ti}$ follow the Gaussian distribution $N(0.9,0.1^{2})$, while Z is assumed to follow the Gaussian distribution $N(0.01,0.01^{2})$. In addition, $N^{+}_{t}$ in all communities follows the Gaussian distribution $N(0.5,0.1^{2})$.\par
\begin{figure}
\centering
\includegraphics[width=1 \columnwidth]{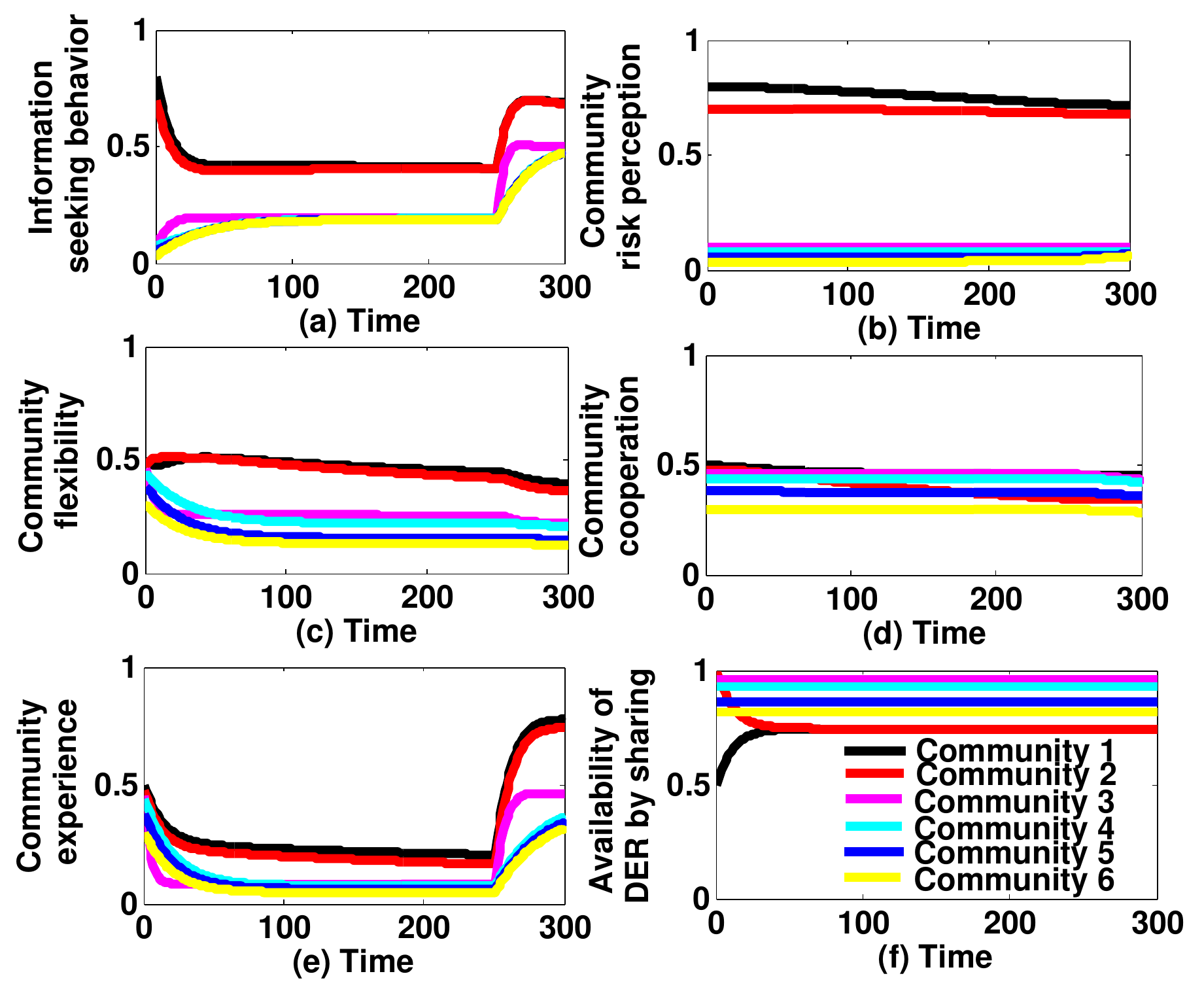}
\setlength{\abovecaptionskip}{-15pt}
\caption{Dynamic change of information-seeking behavior, risk perception, flexibility, cooperation, experience, and the availability of the electricity supplied by DERs for six communities.  The disaster occurs in Community 1. Because Community 2 and 1 are empathetic to each other, the disaster influences the mental characteristics of individuals in Community 2. In addition, other communities are not empathetic at all.  }
\vspace{-0.6cm}
\label{fig:Fig_39}
\end{figure}
\begin{figure}
\centering
\includegraphics[width=1 \columnwidth]{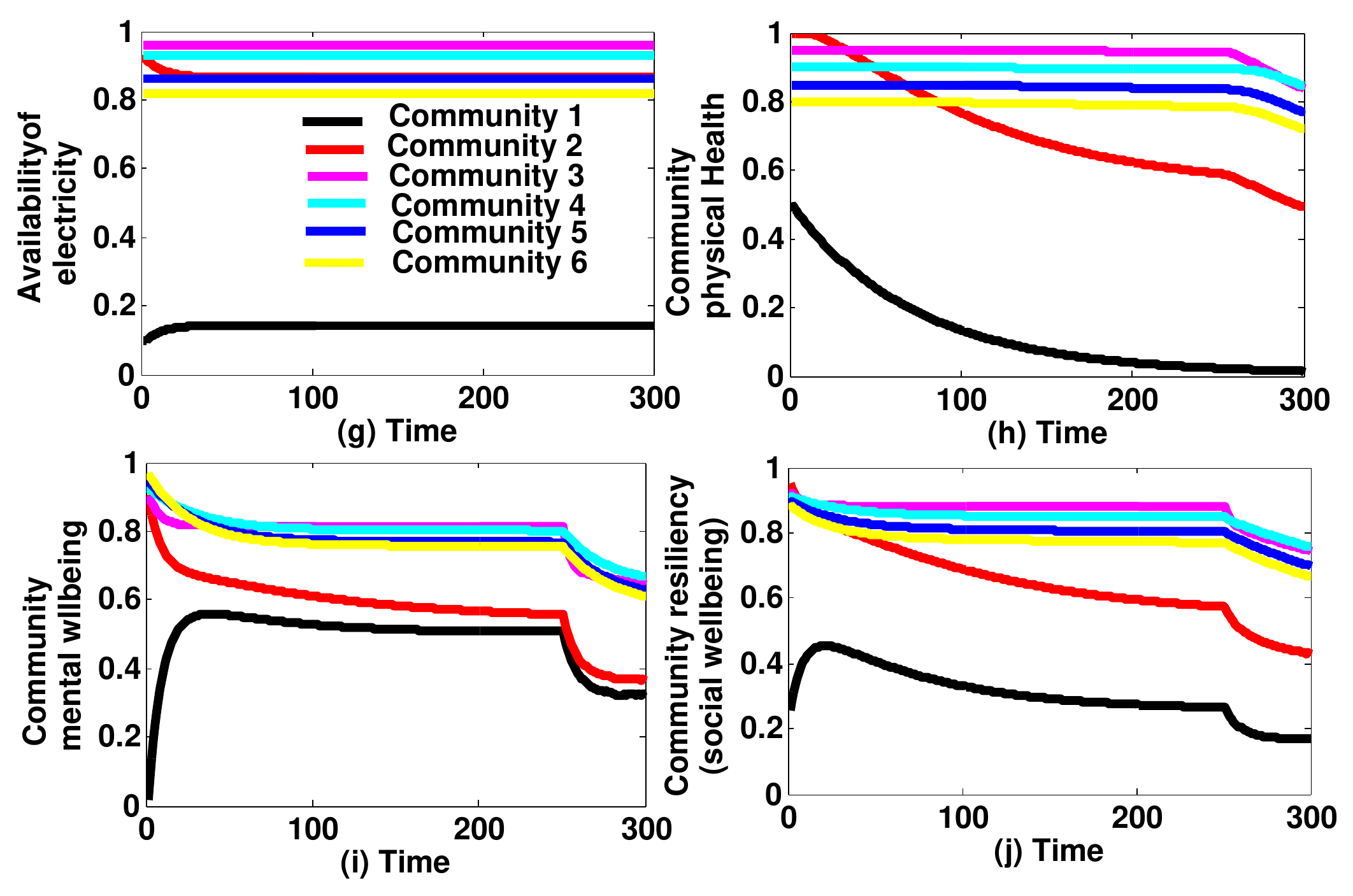}
\setlength{\abovecaptionskip}{-15pt}
\caption{Trends of the availability of electricity, physical health, mental well-being, and community resilience for six communities. Because the disaster happens in Community 1, its people have the lowest level of physical health, mental well-being, and community resilience. The mental well-being of people in Community 2 is affected by this disaster; consequently, community resilience is diminished. }
\label{fig:Fig_39b}
\end{figure}

Figures~\ref{fig:Fig_39} and \ref{fig:Fig_39b} show the average dynamic change of collective behavior and community resilience for the six communities during a disaster. In community 1, because of a high level of the injury factor, the lack of emergency services and electric energy availability from the power grid, a high level of fear and low level of physical health occurs. The level of fear of this community is higher than that of other communities. Because Community 2 has a close relationship with Community 1, their levels of fear are intertwined. As a result, these two communities have a close level of risk perception and information-seeking behavior. Other mental characteristics in these two communities are approximately the same. Community 2 shares its electric energy with Community 1. Hence, the availability of electric energy in the latter is increased. Owing to the fact that the disaster happened in Community 1 and not in Community 2 and due to the higher level of availability of electric energy and emergency services, the physical health of Community 2 is not as endangered as in Community 1. Therefore, people in Community 2  are safe. Furthermore, because of the positive emotion of Community 2 and the high level of empathy between both communities, fear in Community 1 is lowered until time step 2. The feeling of panic among all communities is increased after time step 2 as a result of the mass media, which provides relevant negative news. As a result, the risk perception, the information-seeking behavior, and the experience of the individuals in all communities rise after time step 2. In general, human response in Communities 3 to 6 follows the same trends, resulting in the same status.

\vspace{-0.3cm}
\subsection{Effects of Two Concurrent Disasters on Human Response}

In Example 2, one disaster strikes Community 1, while a second one simultaneously strikes Community 5. The characteristics of Community 1 and its disaster are the same as those of Example 1. The injury factor of disaster in Community 5 is assumed to follow the Gaussian distribution $N(0.1,0.1^{2})$. Electric energy supplied by utilities and emergency services are available. The $M^{E}$, $M^{R}$, and $M^{B}$ of the people in Community 5 follow the Gaussian distribution $N(0.9,0.1^{2})$. Other characteristics of the communities are similar to those of Example 1.  \par

\begin{figure}
\centering
\includegraphics[width=1 \columnwidth]{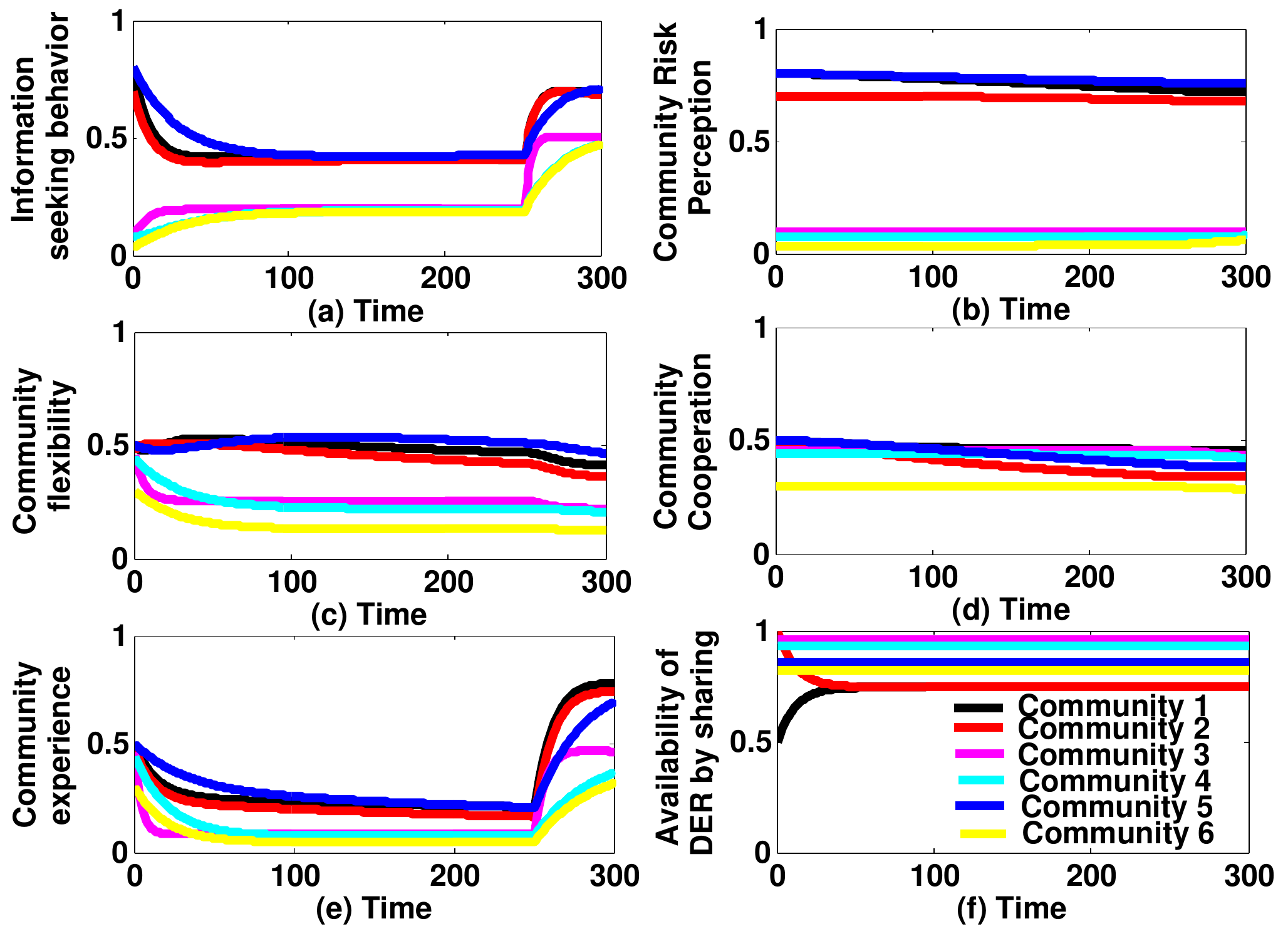}
\setlength{\abovecaptionskip}{-15pt}
\caption{
Dynamic change of information-seeking behavior, risk perception, flexibility, cooperation, experience, and the availability of the electricity supplied by DERs for the six communities. Two disasters occur simultaneously in society. One severe disaster occurs in Community 1. The disaster, which befalls Community 5 is not too dangerous (the injury factor is equal to 0.1).}
\vspace{-0.6cm}
\label{fig:Fig_40a}
\end{figure}

\begin{figure}
\centering
\includegraphics[width=1 \columnwidth]{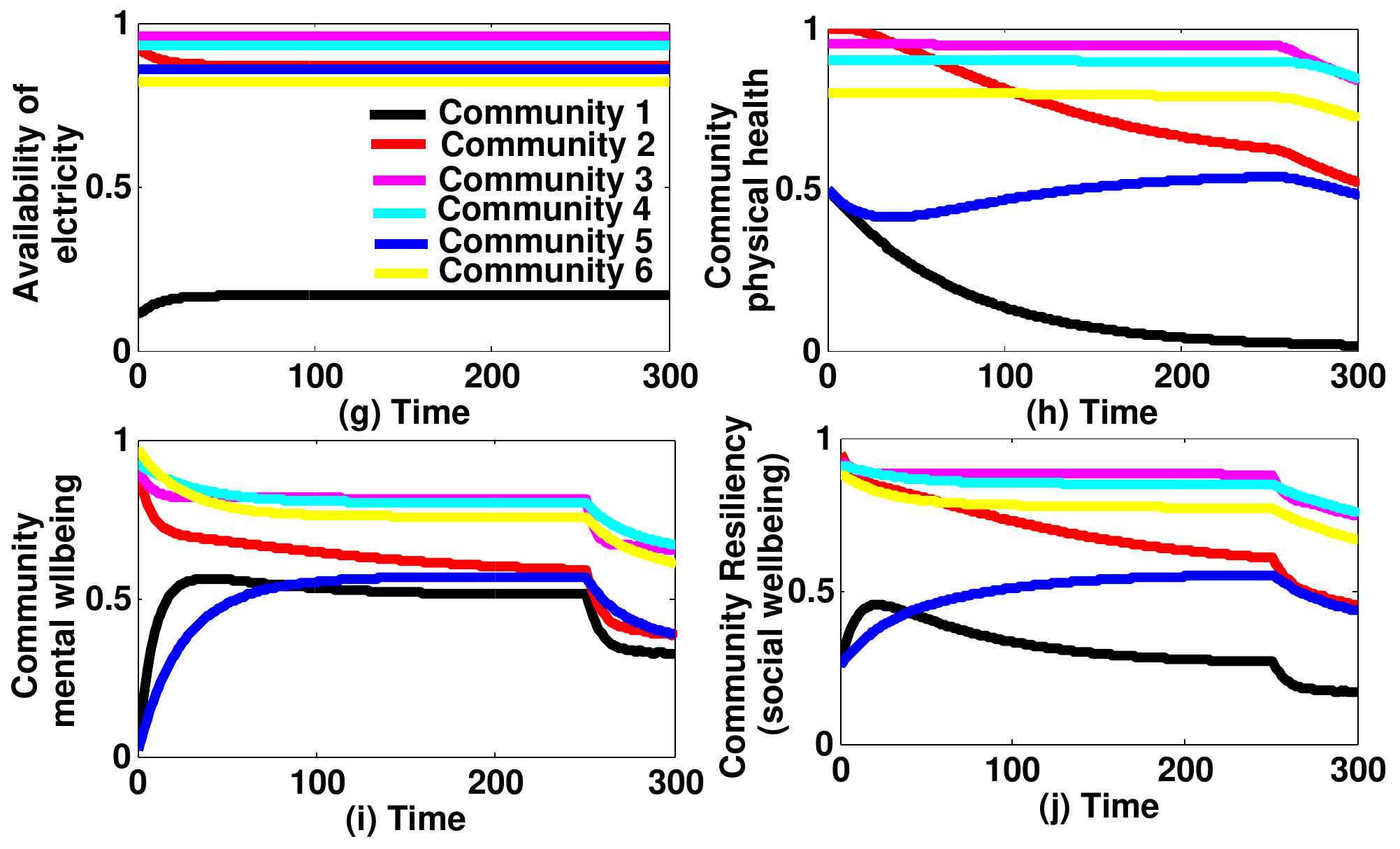}
\setlength{\abovecaptionskip}{-15pt}
\caption{Average dynamic change of availability of electricity, physical health, mental well-being, and community resilience for the six communities.
Although the resilience of Community 1 is similar to that of 5 at the beginning of the disasters, these communities do not have the same trends. Because of the availability of emergency service and electricity during the disaster, the resilience of Community 5  increases over time.}
\vspace{-0.6cm}
\label{fig:Fig_40b}
\end{figure}

Figures ~\ref{fig:Fig_40a} and \ref{fig:Fig_40b} show the average dynamic change of collective behavior and community resilience for the six communities during the disasters. The physical health of the individuals in Community 5 increases because of the availability of power, emergency services, and the low level of the injury factor of the disaster. There is emotion diffusion and empathy among people of Communities 1 and 2. Community 2 does not have any initial panic. People in community 2 are empathetic to Community 1. This is why the level of fear of Community 1 is lower than that of Community 5. Since physical health in Community 5  increases until time step 200, the average level of fear of this community falls.  \par

\vspace{-0.3cm}
\subsection{Effects of the Occurrence of Disasters at Different Times in Separate Communities on Human Response}

In Example 3, one disaster occurs in Community 1 at time step 0, while another occurs in Community 5 at time step 100. The characteristics of Community 1 and its disaster are the same as those of Example 1. The disaster in community 5 causes a power outage at time step 100. All news from the mass media is relevant. The other characteristics of the communities are similar to those of Example 1. 

\begin{figure}
\centering
\includegraphics[width=1 \columnwidth]{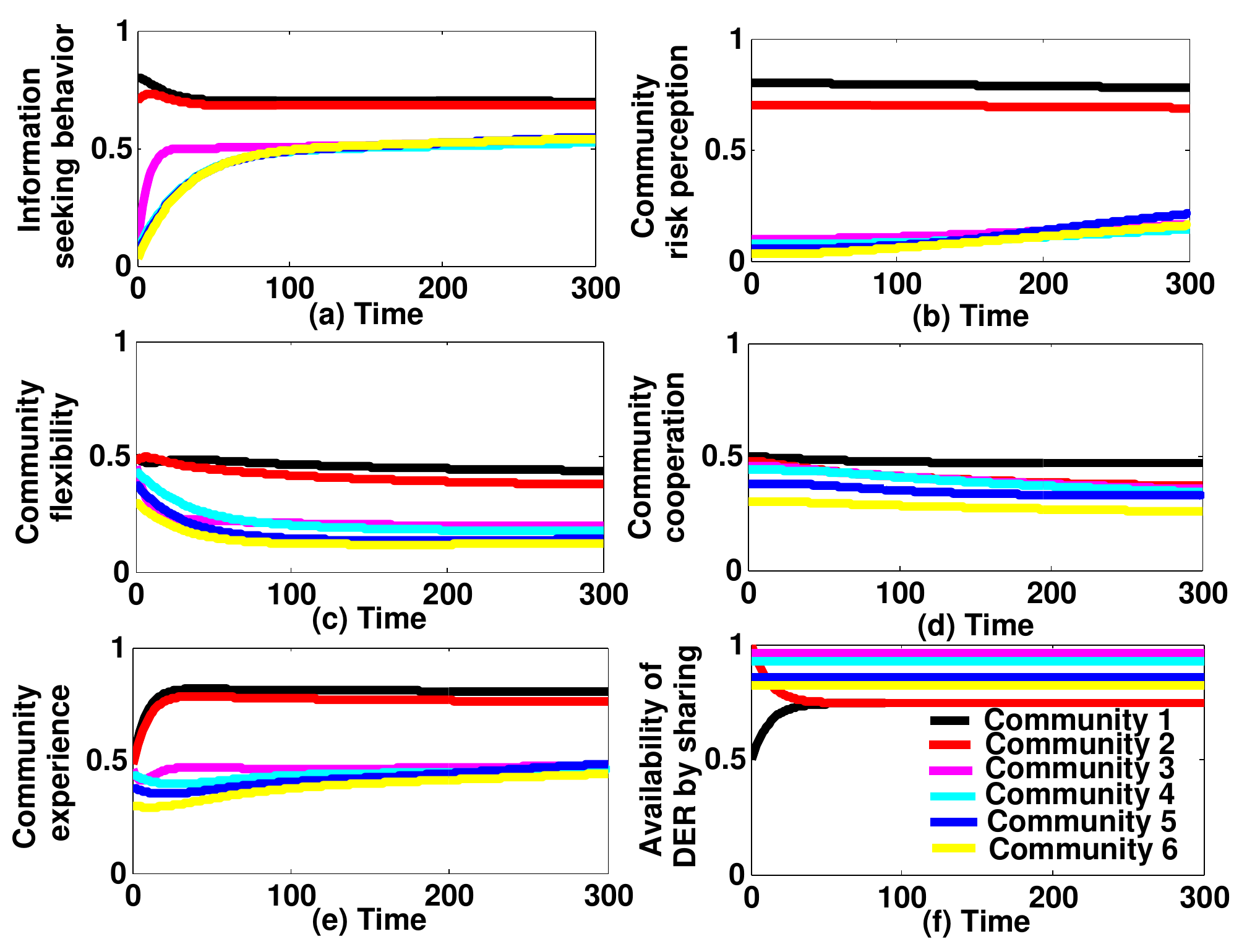}
\setlength{\abovecaptionskip}{-15pt}
\caption{Dynamic change of information-seeking behavior, risk perception, flexibility, cooperation, experience, and the availability of the electricity supplied by DERs for the six communities. Two disasters occur in society at different times. One severe disaster happens in Community 1 at time 0. The other transpires in Community 5 at time step 100. Because of the disaster, the power utilities can not supply electricity to individuals in Community 5.}
\vspace{-0.6cm}
\label{fig:Fig_41a}
\end{figure}

\begin{figure}
\centering
\includegraphics[width=1 \columnwidth]{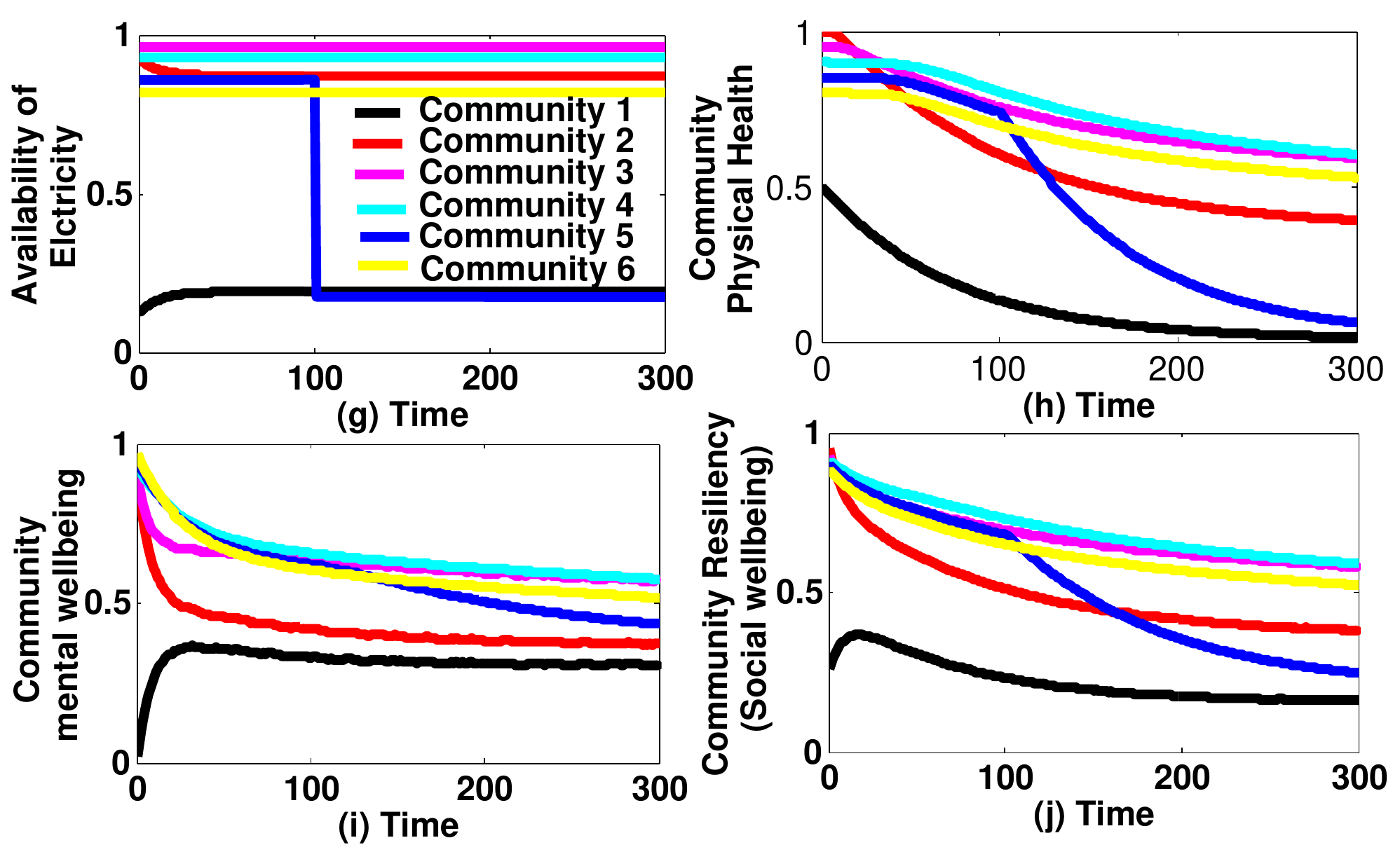}
\setlength{\abovecaptionskip}{-15pt}
\caption{Dynamic change of the availability of electricity, physical health, mental well-being, and community resilience for the six communities. The disaster in Community 5 is very severe. After the disaster occurs in Community 5, the social well-being and community resilience of their people sharply decreases.}
\vspace{-0.6cm}
\label{fig:Fig_41b}
\end{figure}

Figures ~\ref{fig:Fig_41a} and \ref{fig:Fig_41b} show the average dynamic change of collective behavior and community resilience for the six communities during the disasters. As expected, the physical health of Community 5 is sharply lower beginning at time step 100. Understandably, because of the occurrence of the event during this time interval, the level of fear of Community 5 is high.Individuals in this community perceive a high level of risk and seek information. As a result, they obtain experience. 

\vspace{-0.3cm}
\subsection{Effects of Different Population Features on Community Resilience}

\begin{figure}
\centering
\includegraphics[width=1 \columnwidth]{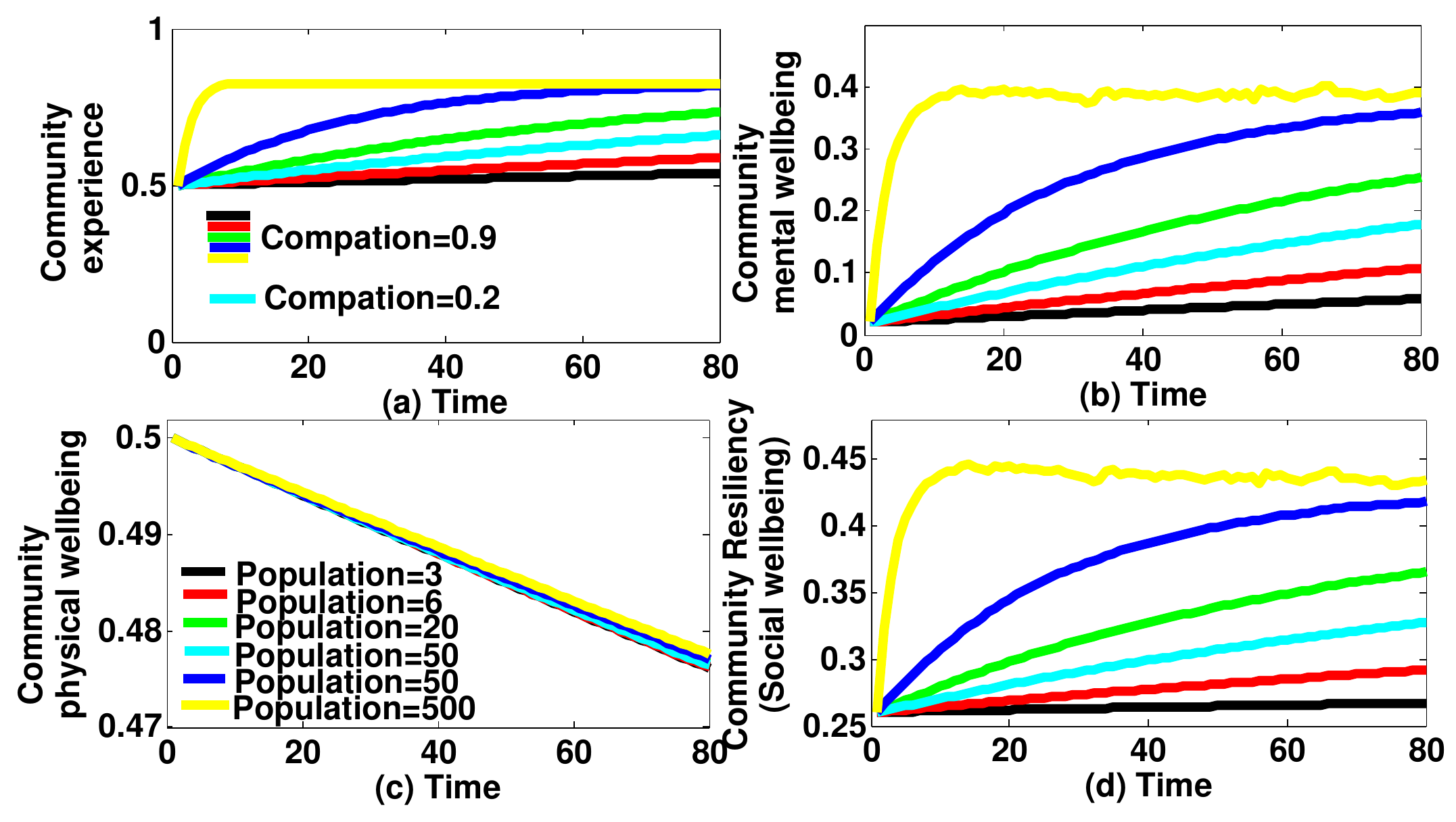}
\setlength{\abovecaptionskip}{-15pt}
\caption{Effects of varying community population on community experience, mental well-being, physical well-being, and social well-being. To signify the importance of compassionate empathy (compassion or empathy), two different values of the empathy with the same community population is assumed. The population of the community is assumed to be 3, 6, 20, 50, and 500. A high population combined with a high level of empathy in the community, can induce a high level of community resilience. }
\vspace{-0.6cm}
\label{fig:Fig_41aa}
\end{figure}

Figure ~\ref{fig:Fig_41aa} shows the changes that occur in the experience, mental well-being, and physical well-being of different population groups who live in a community. This community is similar to Community 1 in that all features, excluding the population size, are the same. An increase in the population size with the same level of empathy is associated with an increase in the level of experience and mental well-being. Moreover, a society with more experience induces a higher physical well-being. If the level of cooperation and experience during and after a disaster is raised, then the level of panic among people is lowered, while the mental well-being is increased. The larger the number of individuals with empathy in a community, the more resilient that community will be. When the population is the same (50), a community with more empathetic individuals (empathy = 0.9) is more resilient  than a community with less empathetic individuals (empathy = 0.2). The relationship among the individuals of a community is an essential characteristic of community resilience. A community with a smaller population (20) and more empathy (0.9) is more resilient than a community with a larger population (60) and less empathy (0.2). In Figure ~\ref{fig:Fig_41aa}, the dynamic changes of other human characteristics are not shown. The fluctuation in mental well-being and community resilience curve is a consequence of the existence of feedback, from other characteristics, i.e., fear, risk perception, information-seeking behavior, cooperation, and flexibility.

\vspace{-0.4cm}
\subsection{Study of the Impact of Disasters on a Community using the EM-DAT International Disaster Database}

Each community has unique attributes, including the injury factor, fear, availability of electric energy, and availability of emergency services. Table~\ref{tab:FCC1} lists the deadliest disasters in the last decade based on an international disaster database. In this table, the features of each disaster are provided.  According to the EM-DAT international disaster database, the injury factor is determined from 0.5+(Death Toll by Disaster Type/ (2* Maximum death)), while the fear from a disaster is determined from 0.5+(Total Number of People Affected by Disaster Type/ (2* Maximum People Affected)). To calculate these factors, average EM-DAT data during 2000-2017 is used. We use real world data\footnote{The data related to the death toll by disaster, maximum death, total number of people affected by disaster type, and maximum people affected are available online at the EM-DAT international disaster database.} as input of the proposed stochastic multi-agent-based model to evaluate the effect of different types of disaster on communities. In addition, we propose how to measure injury factors and induced fear for different types of emergency. 
\begin{table}
\centering
\scriptsize
\caption{Data regarding the injury factor, level of initial fear, and availability of emergency services (AES) and electricity (AE) for different types of disasters based on the EM-DAT international disaster database.}
\begin{tabular}{|p{2.4 cm}|p{1.8 cm}| p{1 cm}|p{.6 cm}|p{0.6 cm}|}
\hline  
Disaster & Injury factor & Fear & AE & AES \\
\hline
Drought              & 0.51473 & 0.83873 & 1 & 0.8  \\
\hline
Earthquake           & 1.00000 & 0.53912 & 0 & 0.4 \\
\hline
Extreme temperature  & 0.61277 & 0.53672 & 0 & 0.5 \\
\hline
Flood                & 0.55873& 1.00000   & 0 & 0.4 \\
\hline
Landslide            & 0.51005 & 0.50152 & 1 & 1   \\ 
\hline
Mass movement (dry)  & 0.50021& 0.50000  & 1 & 1 \\
\hline
Storm                & 0.63776 & 0.69656 & 0 & 0.6 \\
\hline
Volcanic activity    & 0.50033& 0.50097 & 1 & 1 \\
\hline
Wildfire             & 0.50076 & 0.50011 & 1 & 1 \\
\hline
Severe terrorist attack             & 1 & 1 & 0 & 0 \\
\hline
\end{tabular}
\vspace{-0.6cm}
\label{tab:FCC1}
\end{table}

\begin{figure}
\centering
\includegraphics[width=0.8 \columnwidth]{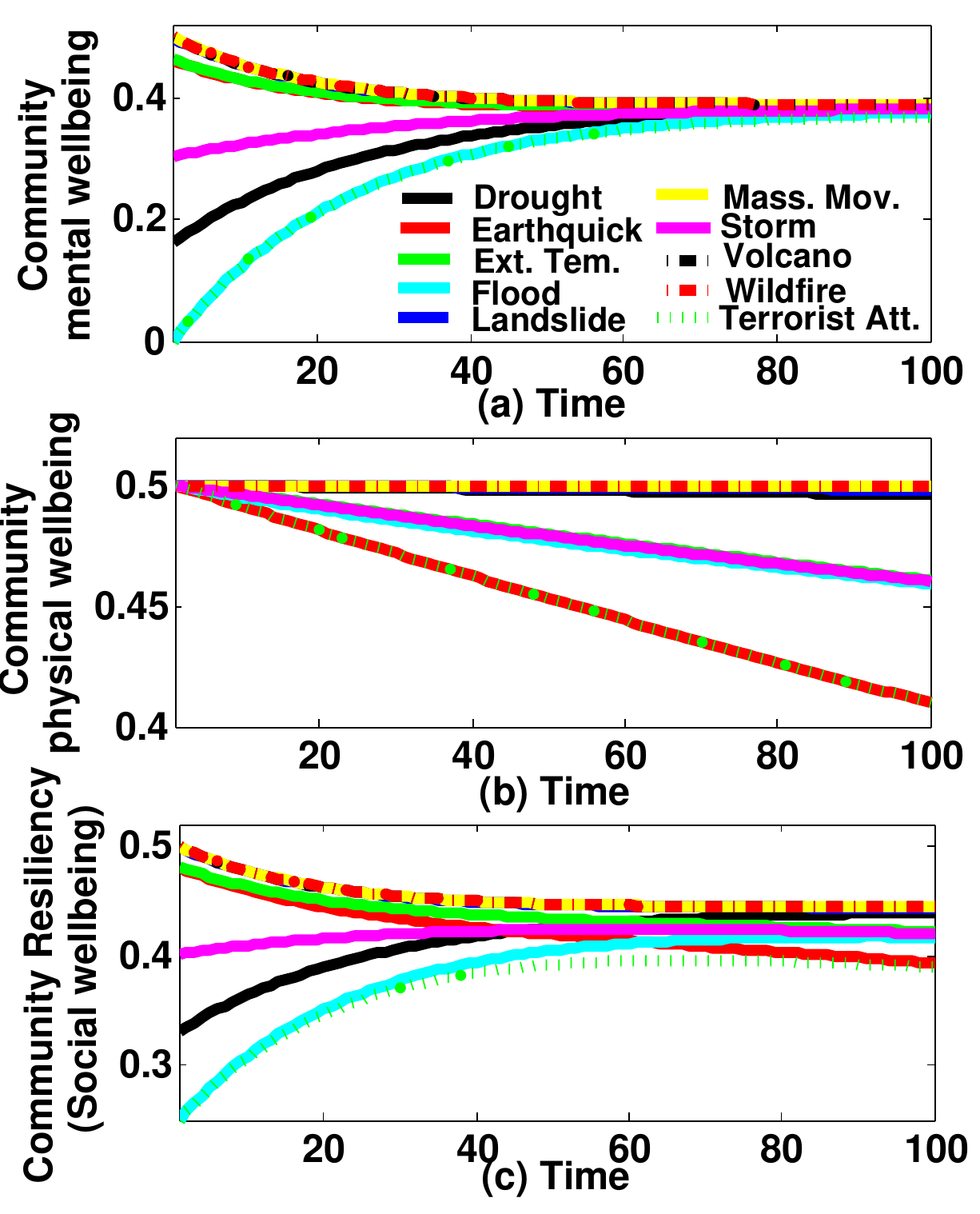}
\setlength{\abovecaptionskip}{-5pt}
\caption{Effects of the most destructive disasters from the EM-DAT database on community mental well-being, physical well-being, and community resilience. These disasters include droughts, earthquakes, extreme temperatures, floods, landslides, mass movements, storms, volcanoes, wildfires, and terrorist attacks (Not from EM-DAT). Each of these disasters is modeled by features like the injury factor, level of initial fear, availability of emergency services, and availability of electricity based on data from the EM-DAT international disaster database. }
\vspace{-0.6cm}
\label{fig:Fig_41bb}
\end{figure}

We consider here a community similar to Community 1 for which Figure ~\ref{fig:Fig_41bb} shows dynamic changes of physical and mental well-being for different kinds of disasters using the EM-DAT database. As seen, 
the community suffering a drought, storm, flood, or terrorist attack shows a low level of community resilience at the beginning of the disaster (between 0 and 0.4). The flood and terrorist attack result in the lowest levels of social well-being (0), while community resilience for all four disasters rises during a given disaster due to the positive human characteristics of community such as cooperation and flexibility.\par 
When the disaster is an earthquake or terrorist attack, the physical well-being of the community sharply drops. In contrast,the physical well-being of the community during storms, extreme temperatures, and flooding only gradually decreases. At the beginning of the disaster, flooding and terrorist attacks induce a high level of fear in the community so that the community's mental well-being is low. This level of panic decreases during a disaster because of an increase in the risk perception, cooperation, and experience of the people. Droughts and storms scare individuals. The other disasters do not induce much fear, and yet the mental well-being of a community during these disasters decreases over time. If the community is more resilient to a specific failure class, it may be more brittle to another failure type \cite{mili2018}. As can be seen from Figure ~\ref{fig:Fig_41bb}, the community is more resilient to some disasters rather than to others.\par

\section{General Idea to Enhance Community Resilience}

This paper has examined the most vital factors of community resilience based on simulations with a proposed agent-based model \cite{jaber2019b}. However, there are some invaluable strategies to enhance community resilience that are not covered in the proposed model:  
\begin{itemize} 
\item \textit{Increase the social well-being of the community during a disaster by time banking:} One course of action to overcome different challenges in the society is alternative currencies, which can lead to community resilience. BERKSHARE in Massachusetts advocates buying local, resulting in greater social well-being for the community \cite{mark2019,wilson2015time}. Other alternative currencies which have been used in past decades include FUREAI KIPPU in Japan in response to elder care \cite{miller2008both,Hayashi2012}; BUS TOKEN in Curitiba, Brazil in response to a garbage problem\cite{macke2018smart,gustafsson2012urban}; TOREKES in Ghent, Belgium in response to high unemployment and urban decay \cite{vandaele2006report}; and TIME BANKING in Blaengarw, Wales in response to events, disasters, and unemployment \cite{boyle2014potential} along with BONUS for emergency situations to help the local economy \cite{moussaid2016patterns,wilson2015time}. Time banking, which is beneficial for community resilience and social well-being, is an adequate alternative currency. Time banking has a local market place so that time instead of money is used as a trading currency. Trade controlled by a dealer is non-reciprocal \cite{ozanne2016alternative, wilson2015time}. All individuals are in control of their community position during the danger. People must not expect the state to rebuild their communities. In this scenario, the presence of human sympathy and cooperation, as mentioned in the article, leads to effective time banking. In addition, although time banking is not against economic banking, it has more impact on improving community resilience during a disaster. Peoples' lives, needless to say, are more essential than cash. Time banking contributes to strong ties among individuals, social support to vulnerable communities, and economic benefit. Economic capacities, communication, social capacities, and community competency are all necessary adaptive facets of time banking. As a result, modeling time banking in the proposed model should be considered to satisfy social well-being during a disaster.  
\item \textit{Other useful concepts for community resilience:} In addition to features like cooperation, experience, flexibility, and empathy, other features including coordination and collaboration can significantly enhance community resilience. Furthermore, there is a difference between cooperation and collaboration. When people cooperate, each individual does one part of the shared aim separately. On the other hand, when people collaborate, there is a direct interaction among people to reach the shared aim \cite{kozar2010towards}.
\end{itemize}

\vspace{-0.2cm}
\section{Conclusions}

This paper provides a general discussion and set of results for two different case studies on the importance of different mental and physical characteristics to the resiliency of individuals and communities during and after disasters. The effect of empathy, experience, flexibility, and cooperation on emotions, risk perception, and information-seeking behavior is investigated. Furthermore, the importance of emergency services, power systems, and DERs is demonstrated. The proposed stochastic multi-agent-based model in this paper is useful for emergent processes and for finding new hypotheses that can be tested in real-world scenarios.The model provides the option of modeling many different effects, which would be costly and difficult to do with only experiments or surveys. The results of this paper provide both agent- and a community-based conclusions. The following conclusions can be extracted from the results of the model:  

\vspace{-0.3cm}
\subsection{Agent-based Conclusions}
The main agent-based results of the proposed stochastic multi-agent-based model are as follow: 

\begin{itemize}

\item When flexibility is high, individuals experience a lower level of panic. Furthermore, the perceived risk of agents is lowered because of the high level of flexibility and low level of fear. As a result, information-seeking behavior which is very much linked with risk perception is diminished. In general, the positive features of individuals may rectify their behavioral drawbacks.

\item  Experience has a negative impact upon the level of fear, information-seeking behavior, and risk perception of agents. It positively influences flexibility if agents are optimistic. When agents do not have previous experience, they seek new information during a perilous situation. Therefore, their experience is increased. There is stable feedback between experience and risk perception in the cognitive process.

\item When emergency services are reduced, the average physical health of individuals falls precipitously. As a result, the level of fear of the agents rises suddenly. Following that, risk perception and information-seeking behavior are also increased.

\item  When the severity of a disaster (i.e., the injury factor) is noticeable, the average physical health of agents dramatically fades.

\item  If all relevant news from the mass media is not very promising, panic increases on average.

\item When the level of cooperation is increased, the agents show a lower level of fear, risk perception, and information-seeking behavior. On the other hand, the feeling of fear during a disaster makes agents cooperate. In fact, a high level of cooperation can positively change individual behavior.

\item When people have a high level of cooperation, they share their electricity sooner than when they have a low level of cooperation. As a consequence, they have a higher level of physical health. Furthermore, due to the high level of cooperation and physical health, people experience a lower level of panic.

\end{itemize}

\vspace{-0.3cm}
\subsection{Community-based Conclusions}
The main community-based conclusions of the proposed stochastic multi-agent-based model are as follow: 
\begin{itemize}

\item The less empathy there is among individuals, the longer other characteristics, including fear, information-seeking behavior, flexibility, cooperation converge to the same level. Additionally, people share their electricity later in the process than when the level of empathy is high. \par
\item When two communities are empathetic to each other and a disaster occurs in one of these communities, the dynamic change of mental characteristics in these two community is roughly the same. \par
\item The higher the population, the more resilient the society is if all individuals have a close relationship with each other.\par

\item The society, whose individuals are closer to each other, has a higher level of community resilience than the community with a lower level of empathy.\par

\item The relationship among the individuals of a community is so vital that the society with less population and more empathy may be more resilient than the community with more population and less empathy.\par
\item If the community is more resilient to a specific failure class, it may be more brittle to another failure type. In other words, the society has a different amount of community resilience under different disasters. A community can be resilient to one disaster while it may not be resilient under other emergencies. Droughts, storms, floods, and terrorist attacks have a low level of community resilience at the beginning of the occurring disaster. \par
\item When the disasters are earthquakes and terrorist attacks, the physical well-being of the community sharply drops. \par

\end{itemize}

\vspace{-0.3cm}
\subsection{Future research}

Although this paper is an important forward step in modeling complex collective behavior for resiliency planning, there are some ideas and challenges, which need to be considered in future work:\par

\begin{itemize}

\item Sentiment analysis can be done to measure mental characteristics based on real data \cite{liu2012sentiment}. The fire hose approach can be used to extract data from Twitter. Moreover, the data obtained by the streaming application interface(API) and search API could be useful. \par

\item It has been suggested to specify the critical electrical loads in each society to enhance community resilience. Supplying critical loads during a disaster is of grave concern. Consequently, there is a need to distinguish among various kinds of loads in this model. Electricity is generated for the consumption of commercial, residential, hospital, and industry loads. In addition, the availability of emergency services and electricity for critical loads must be a priority. One important strategy to enhance community resilience is to ensure electricity continuity for medical services by installing disaster response based hospitals. In General, it is advisable to use disaster response facilities to enhance community resilience.\par

\item The effect of supplying the first 20\% of electricity and critical loads on community resilience is more than the effect of supplying the next 20\% of electricity. This subject must be considered in research associated with community resilience. As a result, the role of DERs on community resilience can be clarified.\par    

\item Each community includes a variety of economic levels of individuals so that they may have a different experience when a disaster occurs. The poor are the most vulnerable. Hence, different economic levels will have different dynamic human responses. Thus, economic level should be included when studying community resilience. \par

\item  In this paper, a simple model of an emergency service network and the power grid is provided. To obtain more reliable results, there is a need to model the real-time operation of the power system and its related technical constraints. The new model can be used to maximize the social well-being of society. Other critical infrastructures can be modeled to study human responses during a disaster. \par 

\item There is a need for a more precise model of disasters. In this paper, features like the injury factor, duration of the disaster, and the initial emotional effect are considered. Other features, such as the mental injury factor, can be considered in the future.  \par  

\item  In this paper, the main concern involves community resilience in which social well-being is a customer-based resiliency metric. Another factor, which can influence resilience, is the time recovery as it relates to the critical infrastructure. In future research, this concept can be added to the proposed stochastic multi-agent-based model. \par

\end{itemize}

\vspace{-0.3cm}


\addcontentsline{toc}{section}{References}
\bibliographystyle{IEEEtran}
\bibliography{neighbourhood}

\end{document}